\documentclass{article}

\usepackage{arxiv}

\usepackage[utf8]{inputenc} 
\usepackage[T1]{fontenc}    
\usepackage{hyperref}       
\usepackage{url}            
\usepackage{booktabs}       
\usepackage{amsfonts}       
\usepackage{eqnarray,amsmath} 
\usepackage{nicefrac}       
\usepackage{microtype}      
\usepackage{lipsum}		
\usepackage{graphicx}
\usepackage[caption = false]{subfig}
\usepackage{float}
\usepackage{xurl}
\usepackage{doi}
\usepackage{natbib}
\setcitestyle{authoryear,open={(},close={)}} 

\usepackage{verbatim}

\usepackage{xr}
\makeatletter
\newcommand*{\addFileDependency}[1]{
  \typeout{(#1)}
  \@addtofilelist{#1}
  \IfFileExists{#1}{}{\typeout{No file #1.}}
}
\makeatother

\newcommand{\degree}[1]{^{\circ}\,\mathrm{#1}}

\title{Extreme rainfall propagation within Boreal Summer Intraseasonal Oscillation modulated by Pacific sea surface temperature}

\date{\today} 					
\author{
    Felix M. Strnad \\
    Cluster of Excellence Machine Learning: New Perspectives for Science,      University of Tübingen, Germany \\
    \AND
    Jakob Schloer \\
    Cluster of Excellence Machine Learning: New Perspectives for Science,
    University of Tübingen, Germany \\
    \AND
    Ruth Geen\\
    School of Geography, Earth and Environmental Sciences,
    University of Birmingham, UK\\
    \AND
    Niklas Boers\\
    Earth System Modelling, School of Engineering \& Design, Technical University of Munich, Germany \\
    Potsdam Istitute for Climate Impact Research, Potsdam, Germany\\
    Department of Mathematics and Global Systems Institute, University of Exeter, Exeter, UK\\
    \AND
    Bedartha Goswami\\
    Cluster of Excellence Machine Learning: New Perspectives for Science,
    University of Tübingen, Germany \\
}

\hypersetup{
pdftitle={BSISO South Asian Monsoon},
pdfsubject={arXiv-SASM-BSISO},
pdfauthor={Felix Strnad},
pdfkeywords={},
} 

\begin{document}
\maketitle
\begin{abstract} 
Intraseasonal variations of the South Asian Summer Monsoon (SASM) contain alternating extreme rainfall (active) and low rainfall (break)  phases impacting agriculture and economies. 
Their timing and spatial location are dominated by the Boreal Summer Intraseasonal Oscillation (BSISO), a quasi-periodic movement of convective precipitation from the equatorial Indian Ocean to the Western Pacific. 
However, observed deviations from the BSISO's canonical north-eastward propagation are poorly understood. 
Utilizing climate networks to characterize how active phases propagate within the SASM domain and using clustering analysis, we reveal three distinct modes of BSISO propagation: north-eastward, eastward-blocked, and stationary. 
We further show that Pacific sea surface temperatures modulate the modes - with El Ni\~no- (La Ni\~na-) like conditions favoring the stationary (eastward-blocked) - by changing local zonal and meridional overturning circulations and the BSISO Kelvin wave component. 
Using these insights, we demonstrate the potential for early warning signals of extreme rainfall until four weeks in advance.
\end{abstract}

\section{Introduction}
The South East Asian subcontinent and its population of around 2.5 billion people receive most of its annual rainfall during the monsoon season from June through September (JJAS) \citep{Webster2020}. 
A defining feature of the South Asian Summer Monsoon (SASM) is the intraseasonal variation of heavy precipitation and convergent wind circulation, which occurs on time scales of $30-60$ days. 
Precipitation peaks are denoted as ``active'' and valleys as ``break'' periods \citep{Wang2018review}; the active phase being often marked by widespread extreme rainfall events (EREs) \citep{Ding2009}.  
The intraseasonal timings of active and break periods can leave long-lasting impacts on crop yields and harvest, and suddenly occurring EREs often wreak havoc on rural and urban infrastructures \citep{Gadgil2006}. 

The Boreal Summer Intraseasonal Oscillation (BSISO) substantially influences the precipitation dynamics over oceans and land masses of the South Asian monsoon domain \citep{Kikuchi2021} and constitutes a major source of rainfall variability on intraseasonal timescales. 
Active phases of the BSISO are initiated in the Indian Ocean, producing easterly wind flows associated with a forced Kelvin wave response, viz. a slowly eastward propagating convective Kelvin wave \citep{Wang1997, Wang2005}.  
As the eastward propagating convective system reaches the Maritime Continent, the convection weakens, and moist Rossby waves are emanated, which then move north-westwards toward India. 
This results in a northwest-southeast tilted band of rainfall that ranges from southern Pakistan in the northwest end to the Philippine Sea and Guam in the southeast. 
The band of heavy rainfall then slowly propagates northward.
One explanation for the northward propagation is proposed by \citet{Wang1997}, according to which the background seasonal mean vertical wind shear interacts with upwardly moving air parcels in the BSISO convective center and, due to the meridional gradient of their vertical velocities, generates cyclonic vorticity and boundary layer convergence to the north of the BSISO cloud band. As a result, the convective system moves northward (cf. \citet{Hoskins2006}). 
The combined movement of the eastward and northward propagation characterizes the ``canonical'' BSISO propagation \citep{Wang2018review}: 
A deep convection zone carrying heavy rainfall emerges in the equatorial Indian Ocean and moves simultaneously eastward and northward, forming a northwest-southeast tilted convection band, and after transgressing the Maritime Continent barrier, progressing further to the Pacific Ocean  \citep{Lee2013, Wang2018review, Kikuchi2021}. 

However, not every anomalous convective activity in the Indian Ocean that is associated with the BSISO follows the canonical propagation pathway. Several studies have reported anomalous convective activity that fails to propagate north-eastward and remains stationary in the equatorial Indian Ocean \citep{Kim2014, Wang2019}.  
One possible factor modulating the variation in the BSISO propagation could be the sea surface temperature (SST) variability associated with the El Ni\~no Southern Oscillation (ENSO), as it is known to affect the rainfall dynamics during the monsoon season, mainly through inducing changes in the Walker circulation \citep{Krishnamurthy2000, Kumar2006}. 
But there has not been sufficient evidence to show clearly that the ENSO influences BSISO propagation and the interactions of the Pacific with the north-eastward propagating convective BSISO system remains poorly understood till date.

Here, we investigate the diversity of BSISO propagation and address the influence of the SST background state by analyzing the occurrence of synchronous EREs over large spatial areas in the SASM domain. 
We use the fact that the BSISO is a large-scale convective system and, thus, BSISO-driven EREs are likely to emerge as spatiotemporally organized weather systems that are connected via long-range teleconnections \citep{Boers2019}.  
We develop a simple heuristic to identify regions of synchronous BSISO-driven EREs by using the framework of climate networks derived from observational rainfall event data \citep{Beck2019}. 
Our method identifies statistically robust geographical regions that tend to have similar active and break phase timings. 
BSISO propagation can thus be investigated as the progression of EREs from one region to the next.  
Based on these propagation pathways, we cluster them by using a spatial clustering method and discover three distinct propagation modes of the BSISO. 
We further find that the background state in the tropical Pacific does affect BSISO propagation but not its initiation in the equatorial Indian Ocean.

While BSISO's impact on annual SASM rainfall has been analyzed thoroughly \citep{DiCapua2020a, Kikuchi2021, Karmakar2021, Hunt2022}, the diversity of BSISO propagation and the potential influence of the SST background state has received less attention. 
Previous studies that have worked on this problem have shown that ENSO can affect BSISO intensity \citep{Lin2008} and propagation over the Maritime Continent. 
In particular, it was found that El Ni\~no-(La Ni\~na-)like conditions suppress (enhance) the propagation over the Maritime Continent  \citep{Liu2016}. 
BSISO activity in the East Asian-western North Pacific region was shown to be influenced by ENSO \citep{Lin2019} and the variability in the northward propagation of the BSISO has been related to different cloud hydrometeors \citep{Abhik2013}. 
Also the east-, north- and north-eastward propagation of BSISO-related convection has been investigated, based on predefined propagation directions \citep{Pillai2016} or on the basis of convective anomalies in the equatorial Indian Ocean \citep{Chen2021}. 
These studies do not find any influence of the background SST state on the propagation and the causes for varying propagation pathways remain unresolved.
Thus, a mechanistic understanding of the propagation diversity is still lacking, limiting the forecast skill of the BSISO \citep{He2019} and the ability of numerical models to describe correctly the north-eastward propagation over the SASM domain \citep{Nakano2019, Kikuchi2021}. 
Our work offers a new perspective on BSISO diversity with implications for improving climate model simulations and shows the potential to develop early-warning signals of EREs along the propagation pathway in a prediction period of more than four weeks in advance.

\section{Results} \label{sec:results}
\subsection{Fingerprint of the BSISO on the spatial organization of EREs}\label{sec:bsiso_communities}

\begin{figure}[!tb]
\centering
\includegraphics[width=1\linewidth]{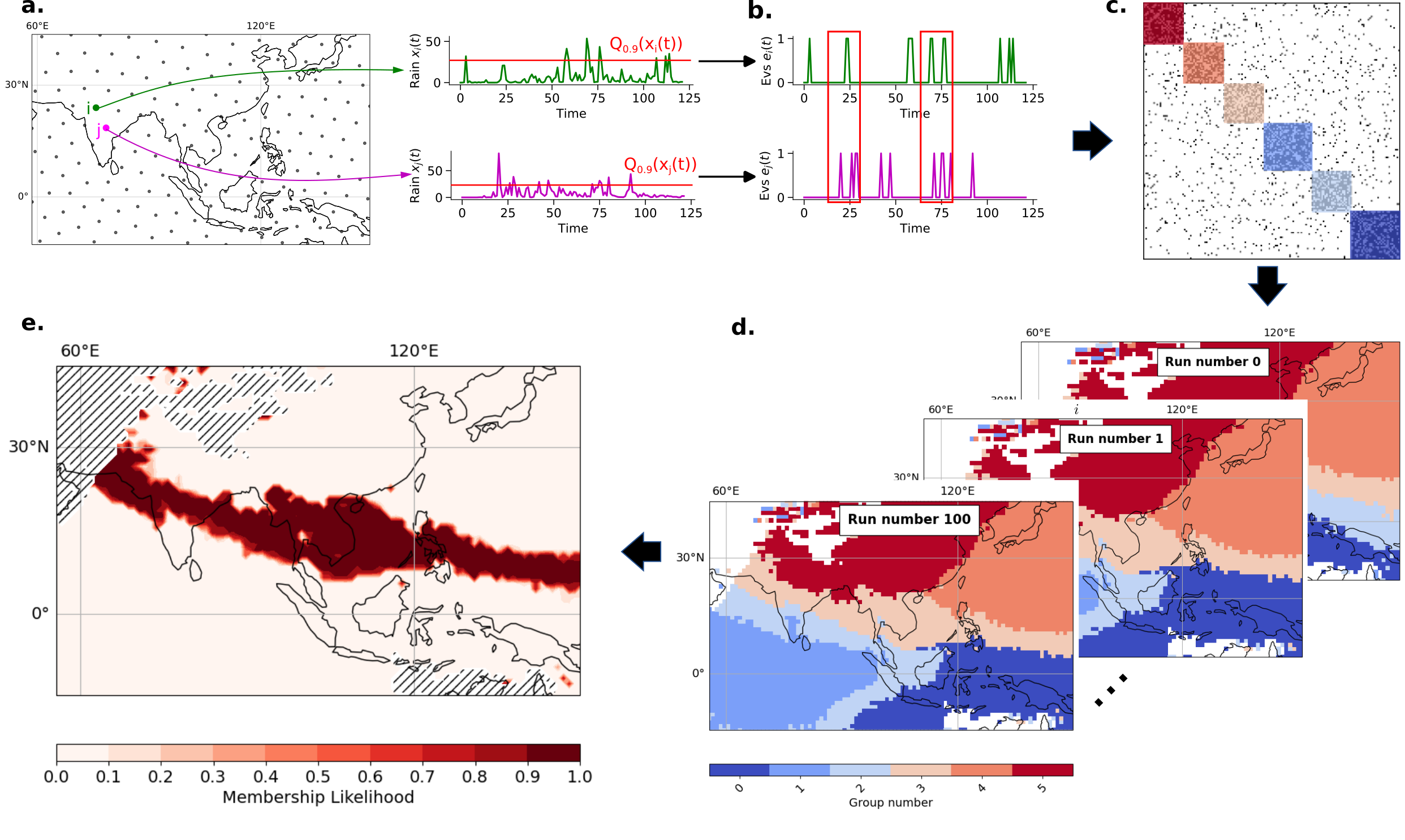}
\caption{\textbf{Identification of communities of synchronous EREs during the SASM.} \textbf{a} The data consists of a set of rainfall time series from spatially uniformly distributed locations (grey circles). 
For visual clarity, only every 5th datapoint is depicted. 
\textbf{b} Rainfall time series are translated into event series estimated locally by days above the 0.9 quantile of all wet days. 
The network is computed based on an event synchronization analysis by comparison of all possible pairs of single event series ($e_i,e_j$) (Material\&Methods Sec.~\ref{sec:es_nets}). 
\textbf{c} The network is represented by its adjacency matrix $\mathbf{A}$, where $A_{ij}=1$ if $i,j$ are statistically significantly synchronous (black dots). The shown matrix is for illustration purposes only.
\textbf{d} Communities are identified as blocks in the adjacency $\mathbf{A}$ and correspond to the spatial locations. 
\textbf{e} A certain community is identified by the overlap of $100$ independent runs of the probabilistic community detection algorithm. The membership likelihood describes the probability of a point belonging to a respective community. Here, one exemplary community is displayed. The additional five communities are shown in Fig.~\ref{fig:msl_all}. Hatches denote regions that are excluded from the analysis.
} 
\label{fig:network_construction}
\end{figure}
In order to explore the BSISO propagation pathway in JJAS, we first detect its signature in regions with similar active and break phase timings. 
We thus identify geographical regions where EREs (defined locally as days with rainfall sums above the 90th percentile of wet days) occur synchronously (within up to 10 days) on average over the full data period, and whose average ERE timings are distinct from the rest of the study area. 
The latter corresponds to ``communities'' of a climate network constructed by estimating event synchronization from extreme rainfall event data of the South Asian monsoon domain (illustrated schematically in Fig.~\ref{fig:network_construction} and explained in detail in Material\&Methods Sec.\,\ref{sec:es_nets} and Sec.\,\ref{sec:community_detection}).  
As our community detection model is inherently probabilistic, we repeat the community detection step several times and use the distribution of different community detection outputs (Fig.~\ref{fig:network_construction}\,c,d) to quantify the membership likelihood of spatial locations of belonging to a particular community (Fig.~\ref{fig:network_construction}\,e and  Material\&Methods Sec.~\ref{sec:community_detection}). 
The low variances in the shape of the communities (Fig.~\ref{fig:msl_all}) confirm that the communities are stable manifestations of synchronous EREs.

\begin{figure}[!tb]
\centering
\includegraphics[width=1\linewidth]{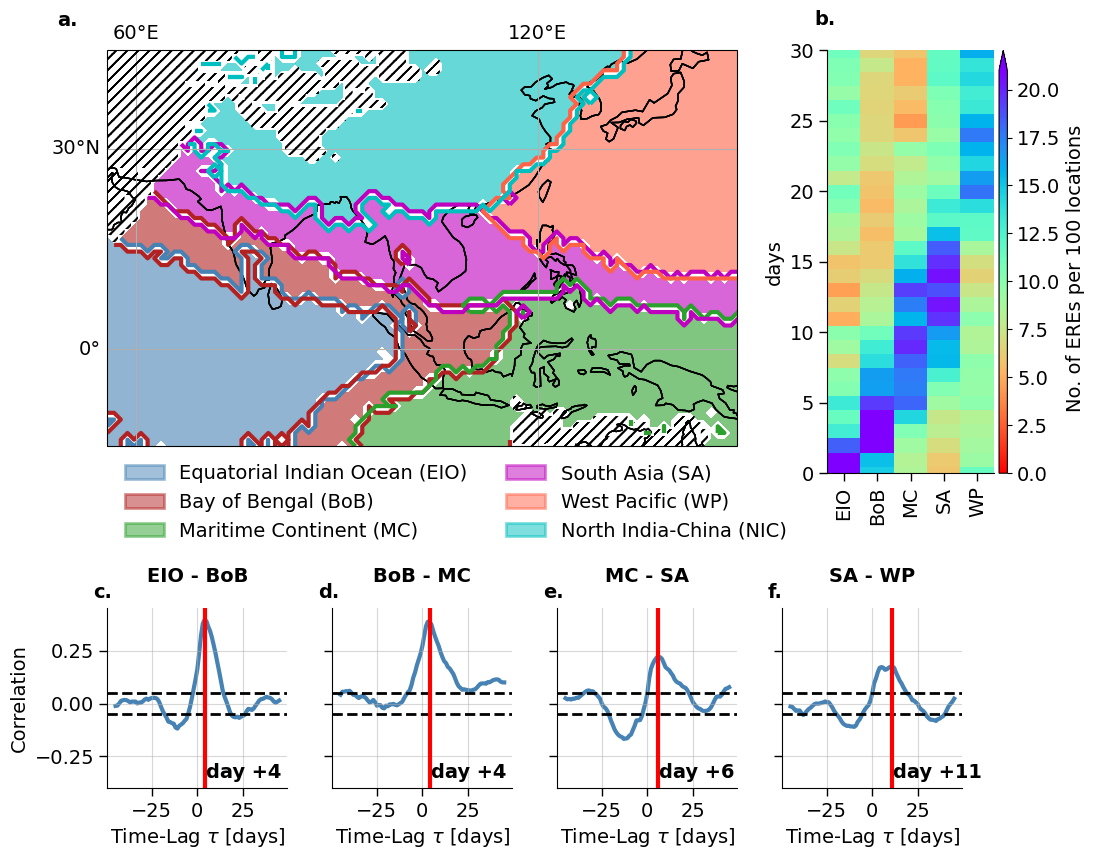}
\caption{\textbf{Communities of synchronous extreme rainfall events during SASM and propagation characteristics.} The communities are determined using a probabilistic community detection algorithm and overlaps of $100$ independent runs. 
\textbf{a} Six regions of synchronous EREs. These are labeled according to their spatial mean position, i.e. equatorial Indian Ocean (EIO), Bay of Bengal (BoB), Maritime Continent (MC), South Asia (SA), Western Pacific (WP), and North India-China (NIC). Hatched areas indicate regions with too little precipitation, which are excluded from the analysis. 
\textbf{b} Temporal evolution of EREs using the most synchronous days within EIO (methods sec.\,\ref{sec:sync_index}), normalized by the number of grid points per community as day $0$. 
\textbf{c-f} Lead-lag correlation analysis between pairs of the synchronous EREs belonging to the communities shown in (a). The time shifts of maximum correlation are denoted by vertical red lines and the respective days are displayed. 
} 
\label{fig:community_detection}
\end{figure}
 
The community detection reveals six geographical regions, labeled here as equatorial Indian Ocean (EIO), Bay of Bengal (BoB), Maritime Continent (MC), South Asia (SA), West Pacific (WP), and North India-China (NIC) (Fig.~\ref{fig:community_detection}\,a).
EIO consists solely of the equatorial and northern Indian Ocean, whereas the BoB region connects both landmasses of India with the Maritime Continent via the Bay of Bengal. 
The V-shaped form of the BoB region has been reported in modeling studies of the BSISO \citep{Wang1997}.  
SA is a northwest-southeast tilted region connecting the South East Asian Monsoon domain with central India also reported in previous BSISO studies \citep{Wang2005, Lee2013, Chen2021}. 
WP is located in the North-Western Pacific north of the Maritime Continent and reveals a long south-north shape along the coastline of East Asia including the island of Japan.  
NIC is solely over land, including the Himalayan Mountains, the Tibetan Plateau, the Ganges Delta, and the Chinese mainland.  
The regions of synchronization show long spatial extension, e.g., the SA spans over approximately $9000\,$km from South Pakistan in the west to the West equatorial Pacific in the east.
The community structure indicates stable synchronization patterns of EREs in both west-east as well as south-north directions.

\paragraph{Propagation of EREs}
During the SASM season, EREs propagate along the sequence of the regions EIO$\rightarrow$BoB$\rightarrow$MC$\rightarrow$SA$\rightarrow$WP, i.e. from south-west to north-east, in approximately 25 days (Fig.~\ref{fig:community_detection}\,b). 
We find that EREs which occur in EIO are particularly likely to take place in BoB +4 days later (Fig.~\ref{fig:community_detection}\,c). 
In the same way, we see that EREs in BoB are likely to arrive at MC at +4 days later (Fig.~\ref{fig:community_detection}\,d), from MC to SA with around +6 days (Fig.~\ref{fig:community_detection}\,e) and from SA to WP at approximately +6\---11 days later (Fig.~\ref{fig:community_detection}\,f). 
EREs within NIC do not show any significant lagged correlation to the other regions, most likely because it is primarily over land, unlike the other regions.  
The propagation sequence of EREs along the identified regions is uncovered by using ``days of maximum synchronization'' within EIO (using the community-specific ERE index, see Material\&Methods Sec.\,\ref{sec:sync_index}), denoted as day $0$, and counting the number of EREs in all communities during these and the following $30$ days. 
The maximum time delay between EREs of different communities is estimated by lead-lag correlation analysis (see Sec.\,\ref{sec:sync_index} and compare Fig.~\ref{fig:sync_ere_index}).

\paragraph{BSISO modulates the ERE occurrences}
\begin{figure}[!tb]
    \centering
    \includegraphics[width=1.\linewidth]{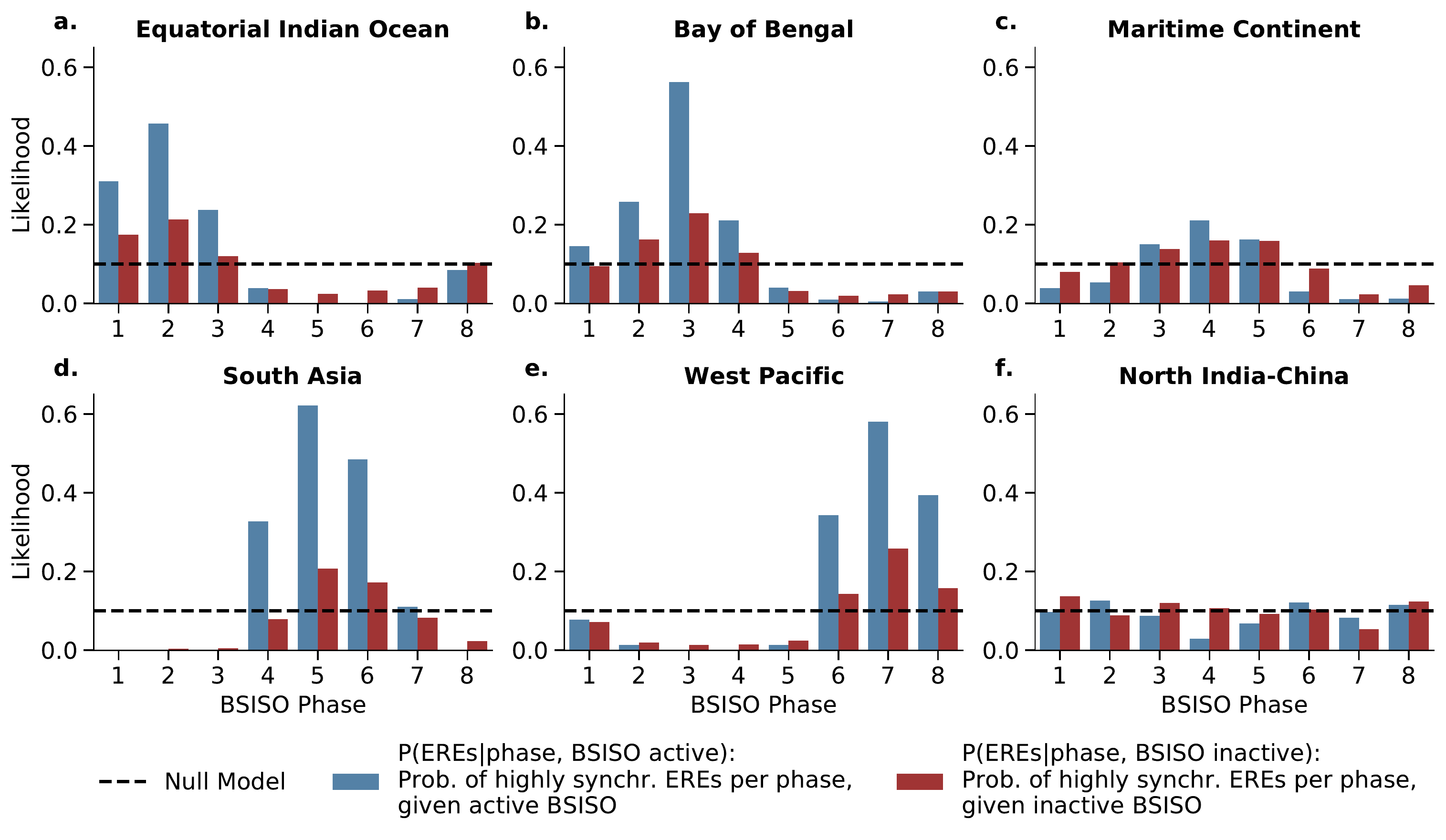}
    \caption{\textbf{Likelihood of synchronous events for active/inactive BSISO phases}. The likelihood of the occurrence of synchronous events ($s=1$) is analyzed for active (blue) and inactive (red) BSISO phases in the regions of Fig.~\ref{fig:community_detection}\,a: 
    \textbf{a} equatorial Indian Ocean, 
    \textbf{b} Bay of Bengal, 
    \textbf{c} Maritime Continent, 
    \textbf{d} South Asia,
    \textbf{e} Western Pacific, and
    \textbf{f} North India-China.
    The dashed line illustrates the likelihood of synchronous events estimated from a null model of randomly distributed synchronous events (i.e. by construction $10\,\%$).   
	}
	\label{fig:bsiso_phases}
\end{figure}
We find that, except for the NIC, in all regions synchronous EREs are significantly more likely to occur in some particular phases of the BSISO (Fig.~\ref{fig:bsiso_phases}). 
We estimate the dependencies between phases of the BSISO (as they are classically defined in \citep{Lee2013}, see also see Material\&Methods Sec.~\ref{sec:data}) and the regions of synchronous EREs (Fig.~\ref{fig:community_detection}\,a) using a conditional dependence test interpreted as the conditional probability of synchronous rainfall events subject to (i) phase (ii) active (inactive) BSISO days (see Material\&Methods Sec.~\ref{sec:cond_indpendence_test}).  
By construction, the null model results in a likelihood of $0.1$ for obtaining synchronous EREs in a certain phase (dashed lines in Fig.~\ref{fig:bsiso_phases}).
For the NIC region, the likelihood of ERES for certain BSISO phases is not considerably different from the null model (Fig.~\ref{fig:bsiso_phases}\,f). 
Therefore, hereafter, we ignore the NIC region and focus on the other five regions (see SI Sec.\,\ref{si:china_india_connection} for a discussion on the NIC).  
While an active BSISO increases the likelihood in the communities substantially, inactive days are distributed close to the null model (Fig.~\ref{fig:bsiso_phases}a-e).  
We observe that the regions that are predominantly oceanic (Fig.~\ref{fig:bsiso_phases}\,a,b,d,e) show a substantially increased likelihood for certain BSISO phases. 
The effect of the Maritime Continent barrier is reflected in the comparably lower likelihood for the MC region (Fig.~\ref{fig:bsiso_phases}\,c) but it still shows an increased likelihood for BSISO phase 4. 
The dependency of the occurrence of EREs and the BSISO can be shown as well be a linear regression test (see SI Sec.\,\ref{si:linear_model}).


\subsection{Modes of BSISO propagation determined by Pacific SST background state} \label{sec:prop_types}
\begin{figure}[!tb]
    \centering
    \includegraphics[width=.85\textwidth]{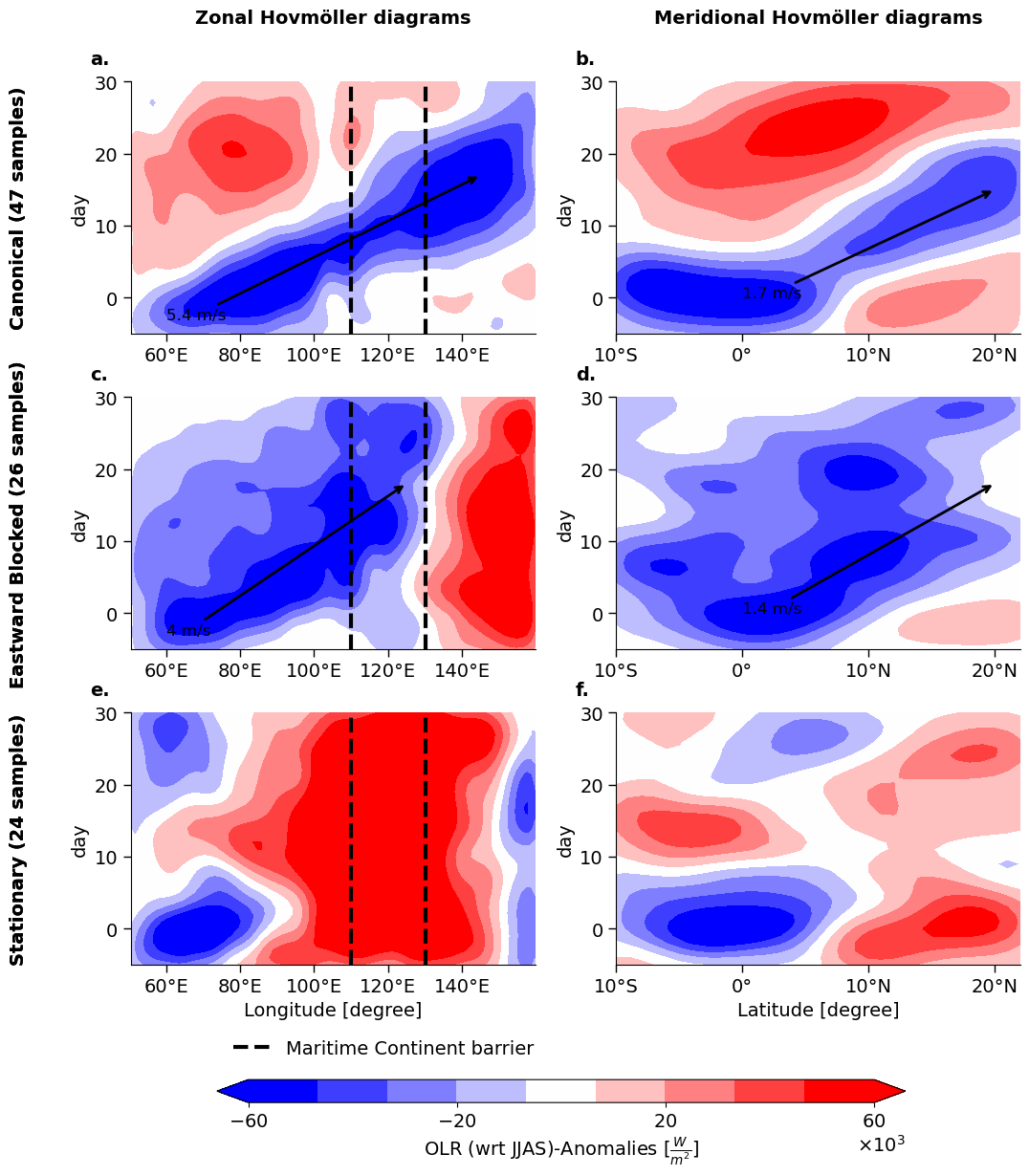}
    \caption{\textbf{Characteristics of the three different BSISO propagation modes.} The first row shows the propagation for the Canonical mode, the second row the Eastward Blocked mode, and the third row the Stationary mode. 
    The first column (\textbf{a,e,i}) shows the composited Hovmöller diagrams in the zonal direction, and the second column (\textbf{b,f,j}) in the meridional direction, for all three modes respectively. 
    All anomalies are computed with respect to the JJAS seasonality. Day $0$ describes the days of maximum synchronization within the EIO community (Fig.~\ref{fig:community_detection}\,a). The dashed lines mark the area of the Maritime Continent barrier roughly estimated to be from $110\degree{E}$-$130\degree{E}$ and the arrows denote the estimated speed of the convective system. 
	}
	\label{fig:propagation_times}
\end{figure}

Since BSISO propagation is clearly accompanied by the ERE progression and the occurrence of regions of highly synchronous rainfalls (Sec.~\ref{sec:bsiso_communities}, Fig.~\ref{fig:community_detection}), we consider the days of maximum synchronization in the EIO region to be potential BSISO initiation time points. 
We create for each individual time point two Hovmöller diagrams of outgoing longwave radiation (OLR) anomalies in the zonal and meridional direction to capture both the east- and northward propagation characteristics of the BSISO. 
We use these diagrams as input samples to a K-means clustering algorithm (see for details Material\&Methods Sec.\,\ref{sec:clustering_prop_times}).  
We obtain three clusters with different propagation features, even though all events initiated in EIO show similar enhanced convection at day $0$:  
\begin{itemize}
    \item \textbf{Canonical propagation} This propagation cluster consists of 47 distinct Hovmöller diagrams. The convective system reveals a propagation speed in eastward direction of approximately $5.4\,$m/s and in the northward direction of approx. $1.7\,$m/s (Fig.~\ref{fig:propagation_times}\,a-b). The phase speed in eastward direction is similar to the speed of the related MJO propagation \citep{Wang2019}.
    The rainfall anomalies travel from the Indian Ocean towards the Western Pacific, passing the Maritime Continent barrier (Fig.~\ref{fig:propagation_times}\,a). 
    The transition over the Maritime continent coincides with the temporary decrease of the convection anomalies at around day $10$. 
    The anomalously wet phase is followed by an anomalously dry phase.  The Canonical propagation mode occurs approximately twice as frequently as each non-canonical propagation mode \citep[cf.][]{Pillai2016}. 
    We confirm that the bipolar pattern between enhanced convection in EIO and dry anomalies at the South Asian mainland and the Western Pacific (Fig.~\ref{fig:propagation_times}\,a) is characteristic in the eastward propagation \citep[cf.][]{Kim2014}.

    \item \textbf{Eastward Blocked propagation}
     This cluster of 26 samples shows a similarly fast northward propagation of $1.4\,$m/s and a suppressed eastward propagation with a speed of $4\,$m/s (Fig.~\ref{fig:propagation_times}\,e-f) that does not transgress the Maritime Continent barrier (dashed lines in Fig.~\ref{fig:propagation_times}\,c).
    Thus, its progression is ``eastward blocked''. 
    Similar to the Canonical cluster, it progresses to latitudes north of $20\degree{N}$ (Fig.~\ref{fig:propagation_times}\,b,f), however, its wet phase is not directly followed by an anomalously dry phase.
    
    \item \textbf{Stationary propagation} 
    This mode, consisting of 24 samples, does not show any propagation. 
    On the contrary, it is even characterized by anomalously dry conditions in zonal and latitudinal directions besides the enhanced rainfalls in EIO (Fig.~\ref{fig:propagation_times}\,i-j).
    Note that keeping the classical definition of the BSISO in mind, these stationary events should not be called BSISO events as the BSISO is defined by the north-eastward propagation. However, there is literature discussing cases of non-propagating EREs that are still associative with the large-scale modes of variability \citep{Kim2014}. 

\end{itemize} 

\begin{figure}[!tb]
    \centering
    \includegraphics[width=.67\linewidth]{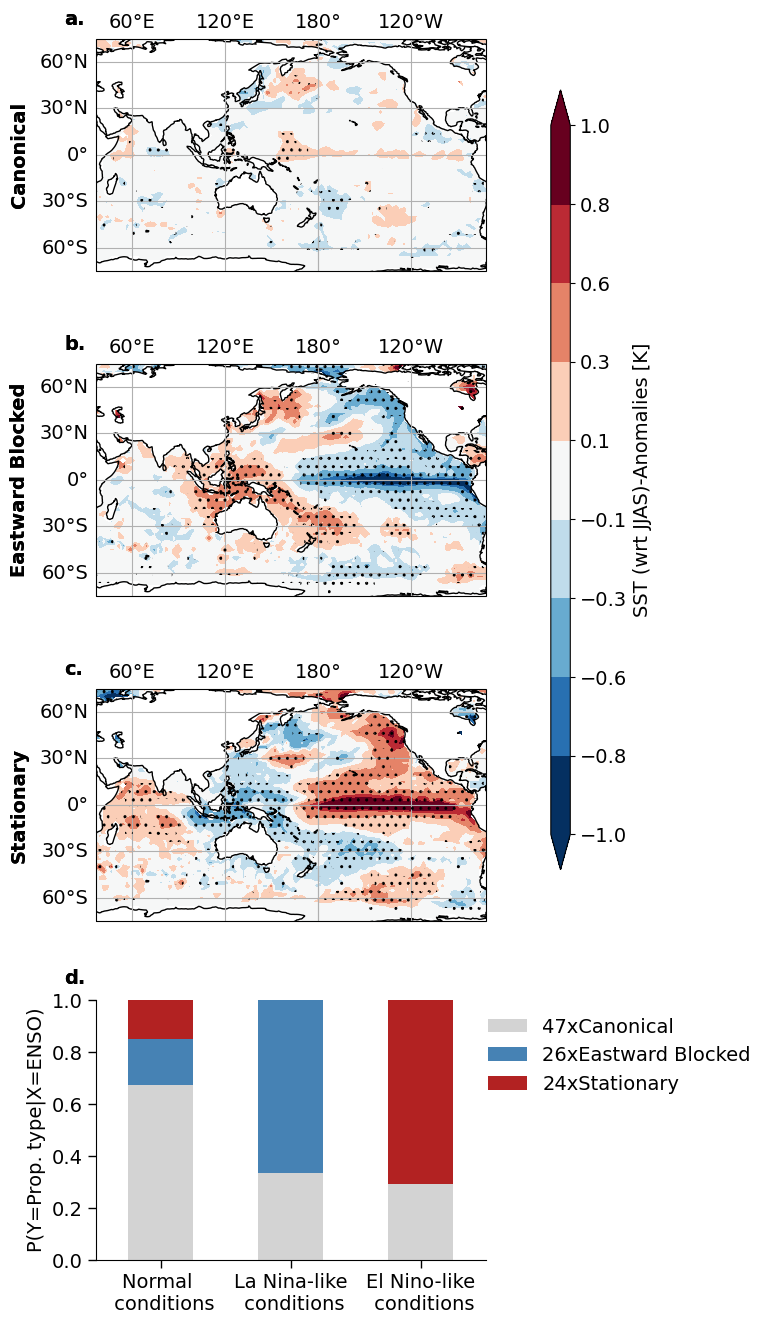}
    \caption{\textbf{SST background state of the three different BSISO propagation modes.}  
    The figure shows the composited anomalies of the background SST conditions at day 0 for the Canonical (\textbf{a}), the Eastward Blocked (\textbf{b}), and the Stationary propagation mode (\textbf{c}). 
    All anomalies are computed with respect to the JJAS climatology, stipples indicate locations that are significant at $95\%$ confidence using Student's t-test. 
    As BSISO propagation times (day 0), we use days of maximum synchronization within the EIO community (Fig.~\ref{fig:community_detection}\,a). 
    \textbf{d} Conditional probabilities for the three propagation modes, given the ENSO condition (La Ni\~na and El Ni\~no based on the NINO3.4 index \citep{Trenberth1997}). The likelihoods for the individual BSISO propagation modes, given the ENSO state, are stacked upon each other.
	}
	\label{fig:sst_background}
\end{figure}
\paragraph{Modulation by SST background state}
We further explore how the Pacific SST background is connected to the BSISO propagation. Fig.~\ref{fig:sst_background} shows the background SST anomalies that are associated with the three BSISO modes.  
We find that the El Ni\~no Southern Oscillation (ENSO) is connected to the deviations of the BSISO propagation from the Canonical mode.
The Canonical propagation mode corresponds to conditions without anomalous SSTs in the Pacific (Fig.~\ref{fig:sst_background}\,a).
A La Ni\~na-like condition, expressed by anomalous cooling in the central Pacific (Fig.~\ref{fig:sst_background}\,b), is likely to occur together with the Eastward Blocked Propagation. 
An El Ni\~no-like condition represented by anomalously warm SSTs in the Pacific Ocean (Fig.~\ref{fig:sst_background}\,c) likely occurs together with the Stationary Propagation.  
We confirm this connection to ENSO using a conditional dependence test (Fig.~\ref{fig:sst_background}\,d). 
The likelihood of a specific BSISO propagation pathway is substantially increased by the respective ENSO state, i.e. Normal conditions favor the Canonical propagation, whereas La Ni\~na- (El Ni\~no-) like conditions favor the Eastward Blocked (Stationary) mode. 
The conditional dependence test is designed to assess the likelihood of a specific BSISO propagation mode subject to the ENSO state (based on the NINO3.4 index definition \citep{Trenberth1997}). 
Still, not all samples match exactly with ENSO (Fig.~\ref{fig:nino34_propagation_modes}). 

\paragraph{Interaction of ENSO with BSISO via the overturning circulation} 
\begin{figure}[!tb]
    \centering
    \includegraphics[width=1.\linewidth]{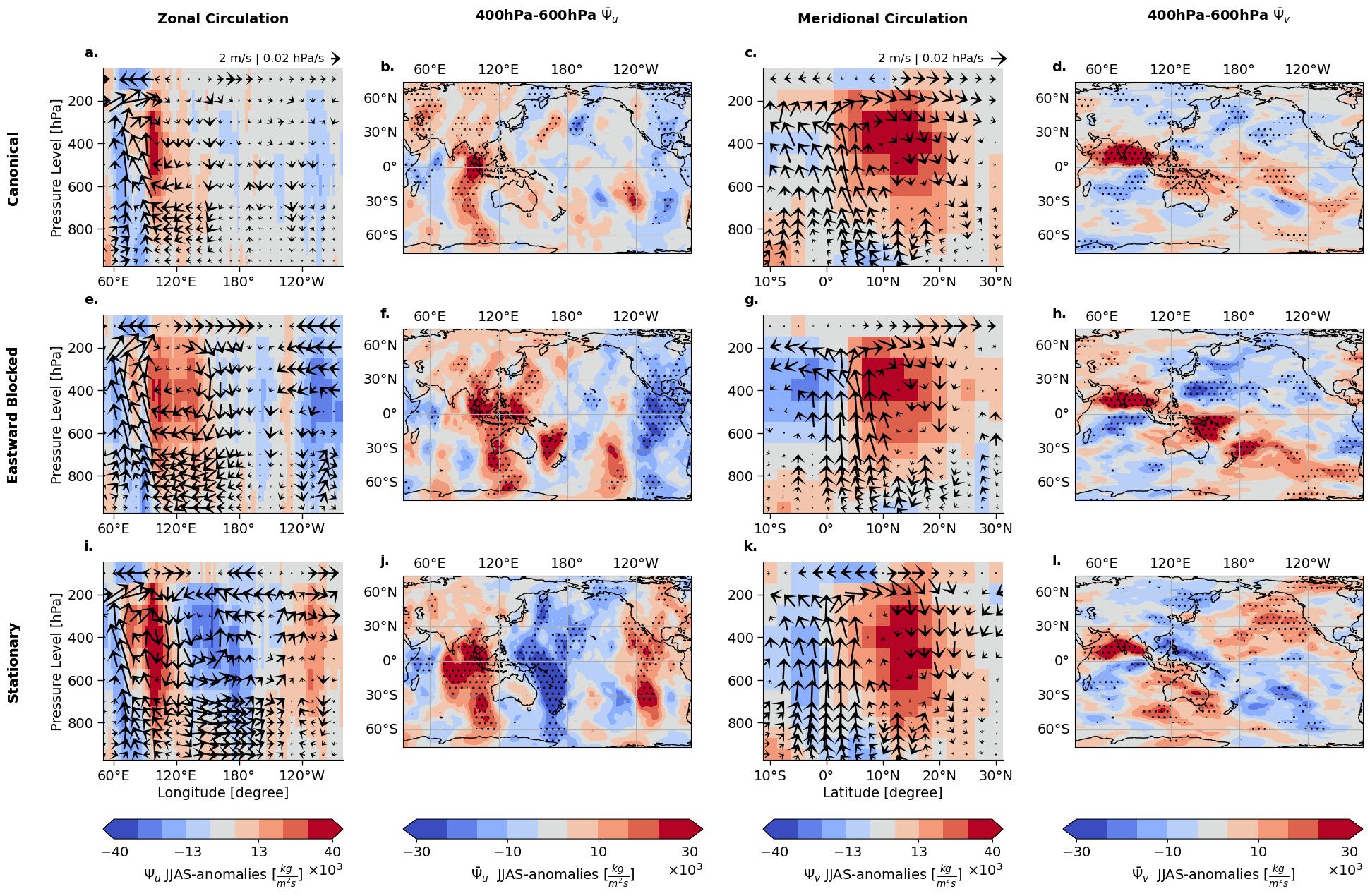}
    \caption{\textbf{Mass stream function anomalies of overturning circulation for ERE propagation modes.} 
    For the different ERE propagation modes (1st-row Canonical, 2nd-row Eastward blocked, 3rd-row Stationary mode) the zonal and meridional overturning circulation anomalies are displayed.
    The first column (\textbf{a,c,e}) shows the composited zonal circulation, meridionally averaged from $0\degree{S}$ to $10\degree{N}$, 
    the second column (\textbf{b,f,j}) the zonally dependent circulation $\Bar{\Psi}_{u}$ averaged between $400$\,hPa and $600$\,hPa. 
    The color shading denotes the mass stream function in zonal direction. Red (blue) indicates irrotational lower-level easterlies (westerlies) and upper-level westerlies (easterlies). 
    The third column (\textbf{c,g,k}) displays the zonally averaged (between $70\degree{E}$-$80\degree{E}$) meridional circulation in the Central Indian Ocean. 
    The fourth column (\textbf{d,h,l}) depicts the meridionally dependent circulation $\Bar{\Psi}_{v}$ averaged between $400$\,hPa and $600$\,hPa. 
    The color shading denotes the mass stream function in the meridional direction. Here, red (blue) indicates irrotational lower-level northerlies (southerlies) and upper-level southerlies (northerlies).
    The wind fields in the zonal (meridional) circulation plots are estimated using the meridionally (zonally) averaged u (v) components, measured in m/s, and the vertical velocity $w$ in the horizontal direction, measured in hPa/s. 
    For visual clarity, only every 3rd wind arrow is plotted.
    Stipples denote anomalies that are significant at a $95\,\%$ confidence level using Student's t-test.  
	}
	\label{fig:vertical_cuts}
\end{figure}
Changes in the local zonal (i.e. the Walker) and meridional (i.e. the Hadley) overturning circulation (Material\&Methods Sec.~\ref{sec:overturning_circulation}) help to understand the interaction of the BSISO with ENSO. 
The Pacific Ocean does not show a significantly anomalous zonal wind flow for the canonical mode (Fig.~\ref{fig:vertical_cuts}\,a), and the enhanced local zonal overturning circulation in the equatorial Indian Ocean (Fig.~\ref{fig:vertical_cuts}\,b) is a consequence of the enhanced convection through the BSISO.
The Eastward Blocked mode reveals an enhanced zonal overturning circulation with the ascending (descending) branch over the Maritime Continent (Fig.~\ref{fig:vertical_cuts}\,e,f).
The induced anomalous Walker cell over the Pacific Ocean with clockwise circulating air masses connects the circulation in the Indian Ocean with the circulation in the Pacific Ocean. 
This circulation deviates from the classical JJAS La Ni\~na condition (Fig.~\ref{fig:vertical_cuts_enso}\,c) in that the Walker cell is shifted towards the West (Fig.~\ref{fig:vertical_cuts_enso}\,c).  
We find two opposing zonal circulations for the Stationary mode (Fig.~\ref{fig:vertical_cuts}\,i). 
The descending branch near the Maritime Continent (Fig.~\ref{fig:vertical_cuts}\,k) separates the anomalous clockwise zonal circulation in the Indian Ocean from the counterclockwise circulation in the Pacific (compare Fig.~\ref{fig:vertical_cuts_enso}\,e). 
We also observe a strongly enhanced Hadley circulation in the Central Indian Ocean for all three propagation modes (Fig.~\ref{fig:vertical_cuts}\,c,g,j) that are the result of the convective anomalies in the EIO. 
This anomalous overturning circulation pattern in the Indian Ocean is primarily driven by the arising convective anomalies of the BSISO in accordance with \citep{Schwendike2021}.  
However, the meridional circulation exhibits strong spatial variations between the different propagation modes (Fig.~\ref{fig:vertical_cuts}\,d,h,i), over the Maritime Continent and the Western Pacific.   
The Eastward Blocked propagation mode shows an elongated convergence corridor over the Western Pacific until the dateline (Fig.~\ref{fig:vertical_cuts}\,h) with an opposing circulation to the circulation in the Indian Ocean.  
The Stationary mode reveals a bipolar pattern in the region of the Maritime Continent and anomalously ascending air around the equator (Fig.~\ref{fig:propagation_times}\,l). 

\subsection{Mechanisms of BSISO propagation diversity} \label{sec:diversity_mechanism}
\begin{figure}[!htb]
    \centering
    \includegraphics[width=1.\linewidth]{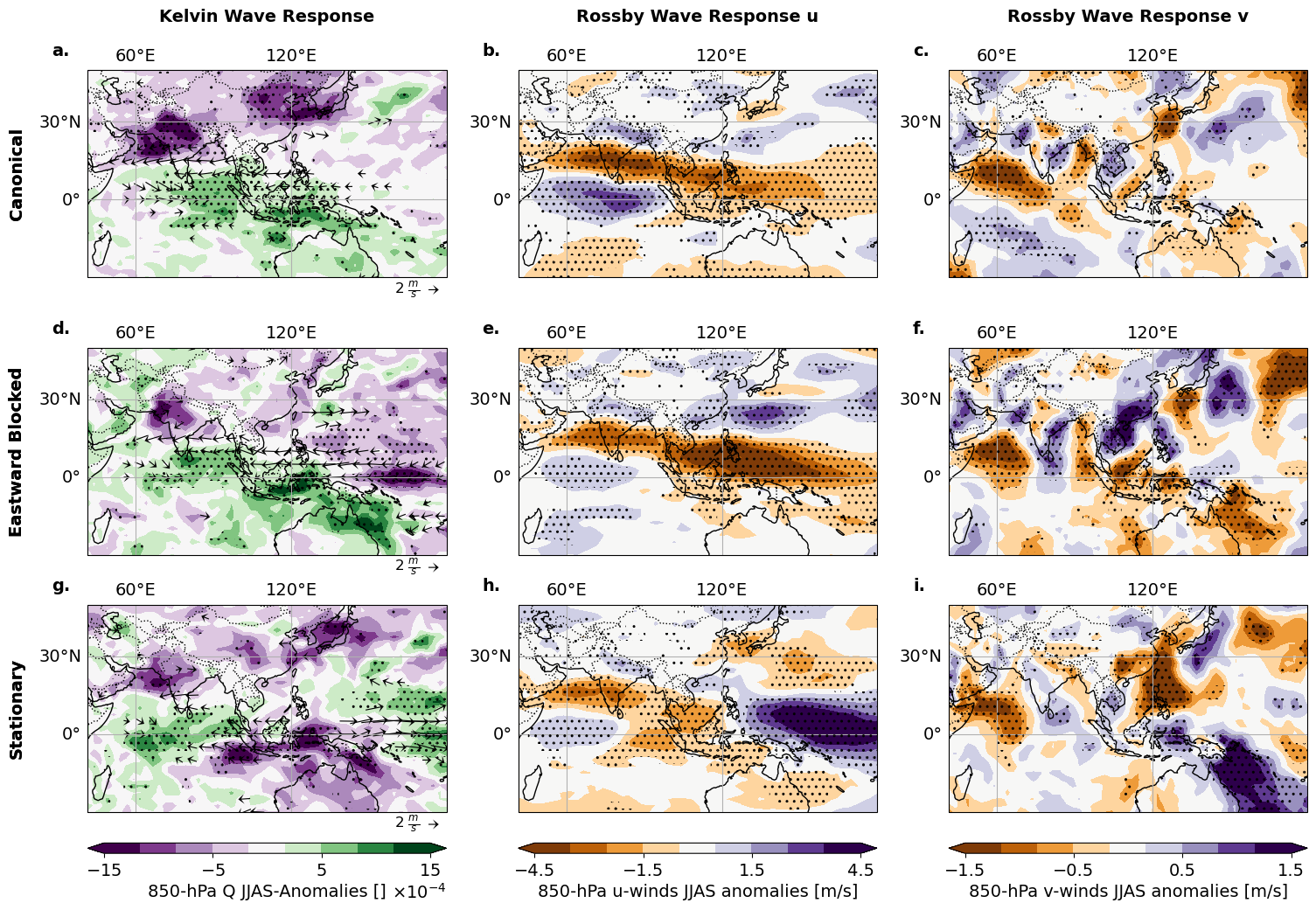}
    \caption{\textbf{Kelvin and Rossby Wave Responses to initiation in EIO.} 
    Different atmospheric conditions as a response to initiation in EIO (Day $0$) are shown for the Canonical (1st row), Eastward Blocked (2nd row), and Stationary case (3rd row). 
    The first column (\textbf{a,d,g}) shows the Kelvin wave response by the mean specific humidity $q$ and mean wind fields at $850$\,hPa plotted as arrows. 
    Only every third arrow of the statistically significant wind field anomaly arrows at $95\%$ confidence level are shown.
    The second column (\textbf{b,e,h}) and third column (\textbf{c,f,i}) show the corresponding Rossby wave response visualized by the zonal wind field component $u$ (2nd column) and meridional wind field component $v$ (3rd column) at $850\,$hPa averaged for $1$ to $5$ days after initiation. 
    In all subplots, stipples denote anomalies that are significant at  $95\,\%$ confidence level using Student's t-test.  
	}
	\label{fig:Kelvin_Rossby_Wave}
\end{figure}
\begin{figure}[!tb]
    \centering
    \includegraphics[width=1.\linewidth]{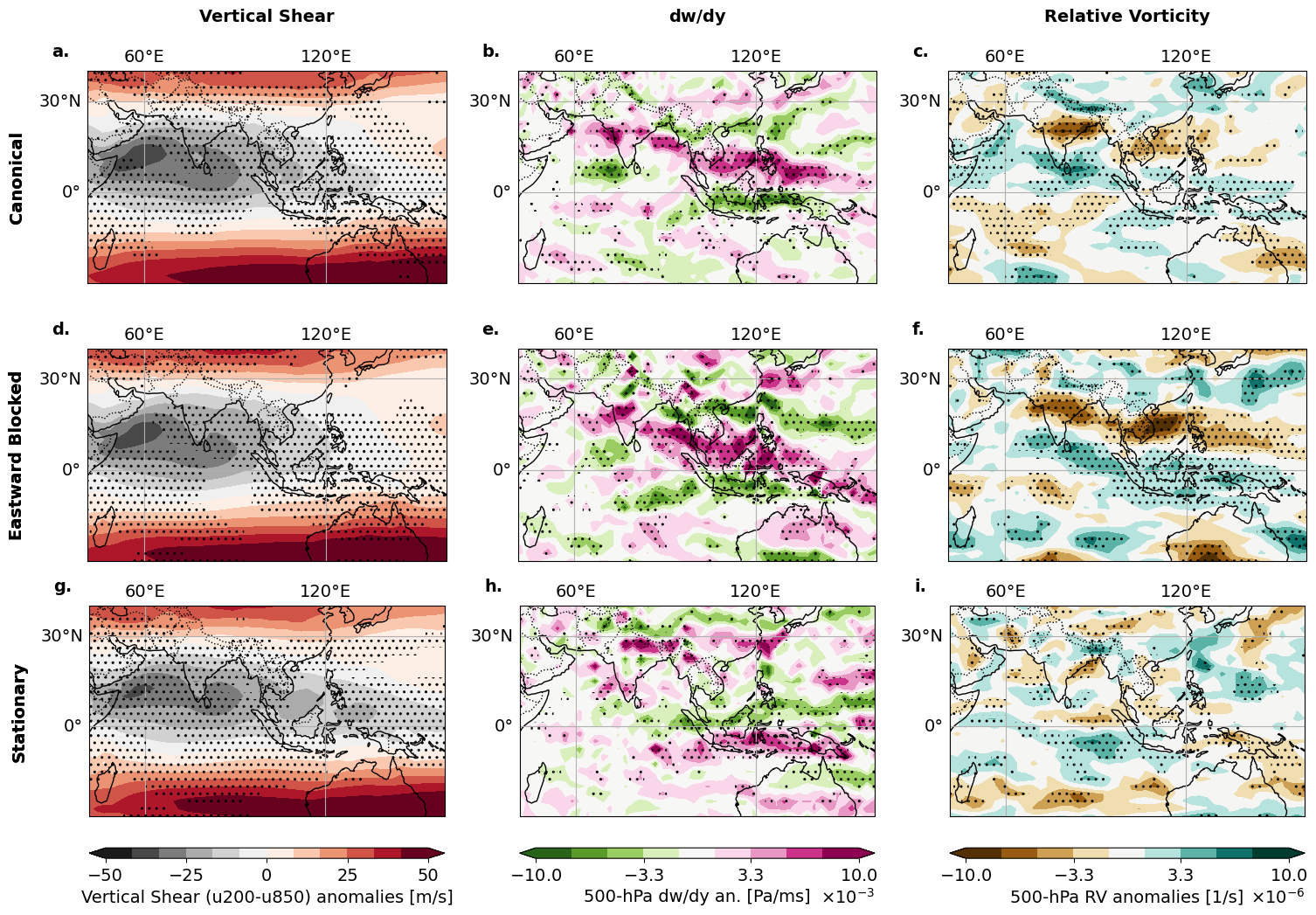}
    \caption{\textbf{Northward propagation for different propagation modes.} The evolution of the northward propagation as a response to eastward propagating Kelvin waves is shown for the Canonical (1st row), Eastward Blocked (2nd row), and Stationary case (3rd row). 
    The first column (\textbf{a,d,g}) shows the asymmetrical mean vertical shear $\Bar{U}_z$, calculated as $u200-u850$ and 
    the second column  (\textbf{b,e,h}) the meridional gradients of the vertical velocity $\frac{\partial w}{\partial y}$ (\textbf{c,f,i}).
    The third column (\textbf{c,f,g}) shows the relative vorticity $\zeta$ as an indicator of the strength of the northward propagation.
    All plots are averaged for days $5$ to $10$ days after initiation. 
    In all subplots, stipples denote anomalies that are significant at a $95\,\%$ confidence level using  Student's t-test.  
	}
	\label{fig:northward_propagation}
\end{figure}

To better understand the modulation of the propagation modes, we put our results into context with the theoretical BSISO propagation mechanism as it is proposed in \citep{Wang1997, Wang2005, Wang2018review, Kikuchi2021}.
We thus analyze the three modes in terms of their Kelvin- and Rossby wave responses (Fig.~\ref{fig:Kelvin_Rossby_Wave}) and their northward propagation mechanism (Fig.~\ref{fig:northward_propagation}).  

\paragraph{BSISO Kelvin wave response}
We argue that differences in the BSISO Kelvin wave component and its coupling to the deep convection affect the BSISO propagation. 
For the Canonical propagation mode, we observe strong Kelvin wave easterlies over the Bay of Bengal to the convection center in EIO (Fig.~\ref{fig:Kelvin_Rossby_Wave}\,a). 
This is consistent with \cite{Chen2019} showing that the substantial north-eastward propagation is the result of a strong Kelvin wave response.
Also, the characteristic moistening preceding the eastward-moving convection center is encountered \citep[cf.][]{Vallis2021}).
We observe that the strength of the Kelvin wave response is reduced for the Eastward Blocked propagation mode (Fig.~\ref{fig:Kelvin_Rossby_Wave}\,d). 
This can be explained by the anomalous warming at the Maritime Continent, leading to ascending air and therefore to a reduced easterly flow over the Bay of Bengal, which in turn results in a weaker Kelvin wave response.
The ascending air at the Maritime Continent also provides an explanation for the eastward blocking (Fig.~\ref{fig:propagation_times}\,d) at around $120\degree{N}$ since the incoming winds from the Pacific Ocean block the eastward propagation of the convective BSISO system.   
The Stationary propagation mode shows a very weak Kelvin wave response (Fig.~\ref{fig:Kelvin_Rossby_Wave}\,g). 
We suggest that the descending dry air at the Maritime Continent leads to winds that are opposite to the convective uprising moist air (Fig.~\ref{fig:Kelvin_Rossby_Wave}\,g). Hence, the Kelvin wave response to the anomalous convection in the EIO fails, so that the deep convection center remains stationary in EIO and vanishes after some days (Fig.~\ref{fig:propagation_times}\,g). 

\paragraph{BSISO Rossby wave response}
The Kelvin wave-guided eastward slow movement of the convective center induces zonally oriented Rossby waves that drift poleward in northeast-southwest tilted bands.  
In the Canonical propagation mode, the Rossby wave response becomes clearly visible in meridional (Fig.~\ref{fig:Kelvin_Rossby_Wave}\,b) as well as in zonal direction (Fig.~\ref{fig:Kelvin_Rossby_Wave}\,c). 
This response also helps to explain the occurrence of the asymmetric V-shaped form of the BoB region (Fig.~\ref{fig:community_detection}\,a) resulting from the decreasing zonal wind speed north- and southward of the equator.  
The Rossby wave response is also present in the Eastward Blocked mode (Fig.~\ref{fig:Kelvin_Rossby_Wave}\,e), although in the meridional (u-) direction the response is slightly weaker. This can be explained by the reduced strength of the Kelvin wave response to the convection in the EIO.  
The corresponding wave train patterns downstream toward the North Pacific (Fig.~\ref{fig:Kelvin_Rossby_Wave}\,f) are similar to the reported MJO teleconnection to the North Pacific, which are in turn regulated by ENSO \citep{Moon2011}.   
The Rossby wave response does not exist in the Stationary mode. 
Even though there is a similar zonal structure (Fig.~\ref{fig:Kelvin_Rossby_Wave}\,f), we do not see an associated wave pattern (Fig.~\ref{fig:Kelvin_Rossby_Wave}\,i), which can be explained by the absence of the eastward traveling Kelvin wave. 

\paragraph{Northward propagation}
The three modes reveal different northward propagation characteristics (Fig.~\ref{fig:vertical_cuts}\,b,d,f) which relate well to the vertical shear mechanism (Material\&Methods Sec.\ref{sec:vertical_shear}). 
The vertical shear is strongest in the northern Indian Ocean for all three propagation modes (Fig.~\ref{fig:northward_propagation}\,a,d,g) and hence the necessary condition for northward propagation is fulfilled.   
The Canonical mode reveals a strong anomalous meridional gradient of the vertical velocity over the Bay of Bengal (Fig.~\ref{fig:northward_propagation}\,b) and north of the Maritime Continent. 
This also explains the anomalous relative vorticity pattern over South India and north of the Maritime Continent (Fig.~\ref{fig:northward_propagation}\,c).   
We find a similar pattern for the Eastward Blocked mode in the meridional gradients (Fig.~\ref{fig:northward_propagation}\,e) and thus also in the relative vorticity (Fig.~\ref{fig:northward_propagation}\,f) with some intensification over the Maritime Continent, which is likely due to the shifted Walker circulation (Fig.~\ref{fig:vertical_cuts}\,f). 
The Stationary propagation mode is not eastward moving and thus not emitting Rossby Waves upon arriving at the Maritime Continent. Therefore, the region in the northern Indian Ocean does not show significant anomalies in the meridional vertical velocity gradient (Fig.~\ref{fig:northward_propagation}\,h) and consequently also no significant relative vorticity (Fig.~\ref{fig:northward_propagation}\,i).

\subsection{Potential for Early-Warning Signals for EREs during SASM}
\begin{figure}[!tb]
    \centering
    \includegraphics[width=.5\linewidth]{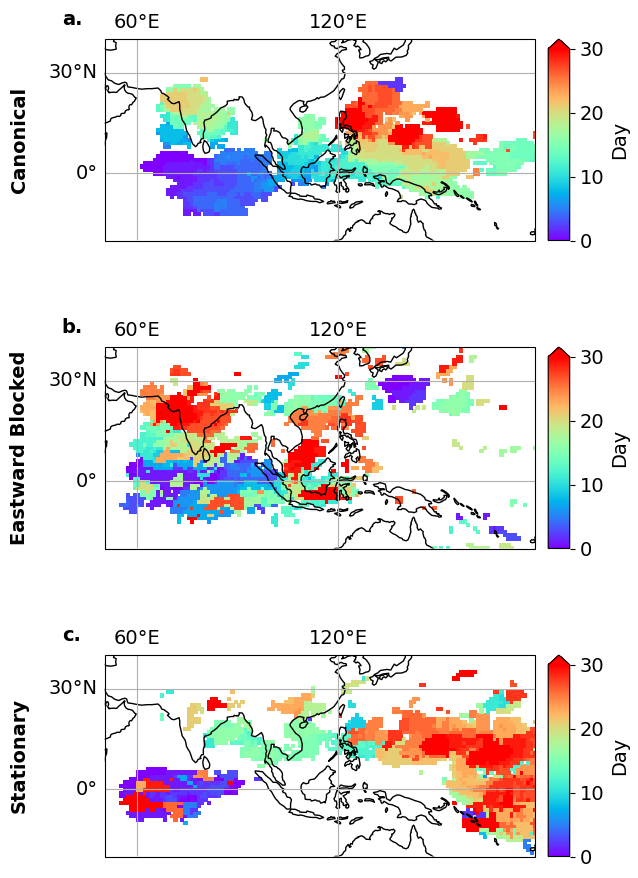}
    \caption{\textbf{Evolution of anomalous rainfall for the three propagation modes.} 
    Regions with the highest rainfall intensity for days after maximum synchronization in the EIO community are shown for the Canonical mode (\textbf{a}), the Eastward Blocked mode (\textbf{b}), and the Stationary mode (\textbf{c}). OLR anomalies are computed according to the JJAS climatology. 
    For every day, the statistically significant mean anomalous OLR above the 95th percentile is shown in its respective color of the day. 
    Day $0$ denotes the days of maximum synchronization in the EIO community. 
	}
	\label{fig:dayprogression_all}
\end{figure}

The uncovered diversity of BSISO propagation has a direct consequence on whether or not a given location in the SASM domain experiences an ERE on a given day once a convective anomaly has been initiated in the EIO. 
It is therefore justified to explore the possibility of Early-Warning signals (EWS) for EREs during SASM that are driven by the BSISO at a time horizon of multiple weeks.  
For normal conditions without substantial SST anomalies in the tropical Pacific, convective anomalies in the EIO are likely to follow the canonical north-eastward propagation (Fig.~\ref{fig:dayprogression_all}\,a, Fig.~\ref{fig:dayprogression_pr_canonical}). 
We sketch the potential for EWS for this mode: We estimate the days of maximum synchronization in all regions and calculate the fraction of events that are subsequent to days of maximum synchronization in EIO within a range of three days. 
In BoB $66\,\%$ of the days of maximum synchronization occur 4\--6 days after days of maximum synchronization in EIO. Subsequently, in the MC region $59\,\%$ of the days of maximum synchronization are observed 6\--9 days later than in the EIO. 
Similarly, $61\,\%$  of the events in the SA community events (within 15\--18 days and $39\,\%$ of the events in WP (within 21\--24 days) are subsequent to days of maximum synchronization in EIO. 
Even this simple approach shows the potential for large-scale spatially resolved ERE predictions up to 25 days in advance which is also the target range of the subseasonal to seasonal (S2S) prediction project \citep{Vitart2017}.

For La Ni\~na-like conditions the propagation is trapped in the region of the Maritime continent, and the heavy rainfall remains in India and the South Asian subcontinent for a longer time span (Fig.~\ref{fig:dayprogression_all}\,b, Fig.~\ref{fig:dayprogression_pr_eastward_blocked}). Moreover, the EREs are likely to propagate towards higher latitudes in northern India (Fig.~\ref{fig:dayprogression_all}).
If El Ni\~no-like SST conditions coincide with a BSISO initiation in the EIO, our results show that India, the Maritime Continent, and the South Asian mainland are likely to only experience very few EREs (Fig.~\ref{fig:dayprogression_all}\,c, Fig.~\ref{fig:dayprogression_pr_stationary}). 


\section{Discussion}\label{sec:discussion}
The aim of this article has been to reveal the specific BSISO propagation pathways and to improve the mechanistic understanding of the BSISO so that a future potential early-warning system for EREs during SASM may be established. 
We uncovered three dynamically resolved propagation modes of the BSISO that show different spatial manifestations and are strongly influenced by the SST background state of the tropical Pacific Ocean. 
We argued that understanding these three modes has important implications for the predictability of the occurrence of EREs in the SASM.  

We demonstrated that the BSISO is a dominant driver of the spatiotemporal organization of EREs during the SASM.
To uncover the BSISO propagation, we introduced a new approach that combines climate networks based on a non-linear event synchronization measure with a probabilistic network community detection algorithm. 
Our approach identified macro-scale structures of spatially coherent patterns of EREs, involving long-range teleconnections between regions from different parts of the SASM domain.  
The results of our community detection approach are stable and consistent also for other community detection implementations \citep{Staudt2014}. 
 Using a posterior likelihood estimation conditioned on the BSISO index, our representation of spatial BSISO locations revealed a skewed distribution over multiple BSISO phases from the classical definition \citep{Lee2013}.  
This confirms the relationship between active and break cycles of monsoon precipitation and the particular phases of the BSISO  \citep[cf.][]{Kikuchi2021}.
Our analysis also provided for the first time a detailed understanding of the spatiotemporal organization of the BSISO-driven rainfall extremes that emerge directly from the data. 
In this sense, our results present an alternative, impact-focused definition of the BSISO based on ERE data which remains still consistent with the classical definition in terms of atmospheric anomalies.

We used the fingerprints of the BSISO propagation to uncover three distinct propagation modes of BSISO-driven EREs: the canonical northeastward propagating mode; the Eastward blocked mode with continuous propagation only in the northward direction while the eastward propagation is intercepted at the Maritime Continent; and the Stationary mode. 
Although the small observational sample size limits the level of statistical significance, our results are robust to randomly chosen initial configurations of the K-means algorithm and subsets of the input data.
Note that a recent study reported three different modes of BSISO propagation as well \citep{Chen2021}, including the ``canonical BSISO'' propagation mode, and an ``Eastward Expansion'' mode, which shares some similarities to the Eastward Blocked mode identified in our work. 
However, \cite{Chen2021} did not find a modulation by the background SST state.
This may be due to differences in the input to their clustering algorithm, which considered data from May--October and used local minima of an Indian Ocean box-averaged intraseasonal OLR timeseries to create pentad mean maps of OLR anomalies. 
Our study focused on the core monsoon season (JJAS) and hence looked at a situation in which the Walker circulation is more shifted to the Western Pacific \citep{Schwendike2014}.  
In addition, we identified BSISO events via regions of highly synchronous EREs ensuring spatial and temporal coherent patterns and accounted explicitly for east- and northward propagation by using Hovmöller diagrams in both zonal and meridional direction to constrain input samples to BSISO propagation characteristics. 

In this study, we identified a plausible mechanism that determines the different propagation modes. 
Our results thus provide a new perspective on how ENSO interacts with the SASM on intraseasonal timescales. 
While the BSISO is known to be marginally correlated with ENSO \citep{Kikuchi2021}, we showed that ENSO state strongly influences the BSISO propagation, leading to local variability in rainfall \citep{Liu2016, Li2019b}.  
We demonstrated how the coupling of ENSO to the BSISO-driven ERE propagation is mediated by the anomalously dry (moist) background moisture and the cooling (warming) near the Maritime Continent during El Ni\~no (La Ni\~na)-like conditions. 
 We further showed how the interplay between the ENSO-related overturning circulation and the BSISO propagation inhibits the east- and northward propagation of EREs under El Ni\~no-like conditions.
 This is in agreement with ISM failures observed during El Ni\~no events \citep{Kumar2006, Joseph2011}.  
Conversely, La Ni\~na-like conditions in the tropical Pacific are found to favor the Eastward Blocked mode, such that EREs move only northward and  bring extended periods of strong rainfall to the South Asian and Indian mainland as well as to the Maritime Continent.
This result further ties in with the longer active spells observed in La Ni\~na years over Indian and South Asian mainland \citep[cf.][]{Joseph2011, Dwivedi2015}.
For El Ni\~no conditions the convection is suppressed, whereas for La Ni\~na conditions the anomalous warming at the Maritime Continent leads to opposing winds that prevent the propagation beyond the Maritime Continent. 
Consequently, the Maritime Continent experiences a weakening (enhancement) of the rainfalls for El Ni\~no (La Ni\~na) conditions corroborating results of \citep{Liu2016}. 
Our results thus imply a pronounced role of the geographical position of the Maritime Continent region for the BSISO propagation and offer a new perspective on its barrier effect upon existing theories based on the complex topography \citep{Kim2017}, the high land-sea thermal contrast \citep{Zhang2017}, and the diurnal convection over land \citep{Ling2019}.  

The proposed mechanism of BSISO diversity may provide a framework for understanding why models fail to simulate the BSISO propagation over the SASM domain realistically \citep{He2019, Nakano2019, Kikuchi2021}, and thus could offer a new validation scheme for GCM development.
In addition, we outlined the potential for medium-range forecasts of EREs and for developing a risk assessment for floods in the South Asian Monsoon domain and along the coast of South East Asia \citep{Zhu2003, Li2015}, on the scale of more than 4 weeks in advance. In comparison, current forecast models of the European Centre for Medium-Range Weather Forecasts (ECMWF) are predicting at a forecast lead time of around 14 days \citep{Wu2022}.
Although the discussed Early Warning Signals approach is simplistic, the prediction skill could likely be substantially increased and a location-specific early warning indicator could be developed by using tools from pattern recognition tasks in machine learning in a similar way as it has been proposed in a recent perspectives paper \citep{Mariotti2020}.

\section{Materials and Methods} \label{sec:methods}
\subsection{Data} \label{sec:data}
We use daily precipitation sums for the time period of 1979---2020 from the MSWEP dataset \citep{Beck2019} that merges gauge, satellite, and reanalysis data provided in a resolution of $0.1^{\circ}$x$0.1^{\circ}$. We restrict our analysis to the South Asian Monsoon domain ($55\degree{E}$ \--- $140\degree{E}$, $20\degree{S}$ \--- $50\degree{N}$). We use next-neighbor interpolation to map the data to a grid of spatially approximately uniformly distributed points employing the Fekete algorithm \citep{Bendito2007}. The distance between two points corresponds to the spatial distance between two points at the equator of a Gaussian $1^{\circ}$ grid, resulting in a total of $\approx 4700$ grid points. 

We linearly detrend the precipitation time series. The event time series is constructed from `wet days' only, defined as days with rainfall of at least 1 mm/day. ERE days for a single location are defined as those days where the daily precipitation sum exceeds the $90$th percentile of all wet days.  
Climatologies on days with high ERE synchronicity, are computed for top net thermal radiation (translating to Outgoing Longwave Radiation), sea surface temperature, and multi-pressure level variables on $50$--$1000$\,hPa of $(u,v)$-wind fields, vertical velocity $w$, and specific humidity $q$ taken from the ERA5 Global Reanalysis dataset \citep{Hersbach2020}. The datasets are interpolated to  $2.5^{\circ}$x$2.5^{\circ}$ grid.  
The NINO3.4 index was estimated according to \citep{Trenberth1997} using SST anomaly fields from the ERA5 dataset. 

The daily resolved BSISO index by \cite{Lee2013} is taken from \url{https://apcc21.org/ser/moni.do?lang=en} (Last Accessed: 14th June 2022). 
The index is calculated by multivariate Principal Component Analysis of OLR and near-surface u850\,hPa, v850\,hPa winds of the region $10\degree{S}$ \--- $40\degree{N}$, $40\degree{E}$ \--- $160\degree{E}$ for days from May until October \citep{Lee2013}. 
OLR is a useful indicator for deep convective activity and the wind fields reflect the north- and eastward movement \citep{Waliser1993}.  
The first two leading principal components (PCs) are used to define the state of the BSISO by the amplitude $A=\sqrt{PC_1^2 + PC_2^2}$, where $A\geq 1.5$ ($A<1.5$) is called active (inactive).  
The two-dimensional space spanned by $PC_1$ and $PC_2$ is subdivided into eight equally sized sections that denote the phase of the BSISO.   
There are further BSISO indices available (e.g. \cite{Kikuchi2012, Kiladis2014}) but the qualitative differences are minor \citep{Wang2018}. 

\subsection{Event synchronization-based climate networks} \label{sec:es_nets}
Assume a spatiotemporal dataset $X\in \mathbf{R}^{N\times T}$, where $N$ denotes the number of datapoints and $T$ is the number of points in time. 
The climate network $\mathcal{G}$ is defined as $\mathcal{G}=(V,E)$ where each geographical position of the dataset $x_i(t) \in X$ corresponds to a node $n\in V$ and $E$ is the set of edges. 
Network edges $e_{ij}\in E $ encode strong statistical dependencies between pairs of time series $x_i(t)$ and $x_j(t)$. 

\paragraph{Event Synchronization.} The number of temporally coinciding events is counted between pairs of event sequences $\{e^m_i\}_{m=1}^{s_i}$ and $\{e^n_j\}_{n=1}^{s_j}$, where $s_i$  ($s_j$) are the total number of events at location $i$ ($j$), and $e^m_i$ ($e^n_j$) describes the timing of an event in $i$ ($j$). 
The delay between an event $e^m_i$ in $i $ and an event $e^n_j$ in $j$ is denoted as $d_{ij}^{m, n} = e_i^m - e_j^n$. 
Defining the set $D_{ij}(e^m_i, e^n_j)$ as the set that contains all four neighboring events of $e^m_i,e^n_j$,

\begin{equation}
    D_{i,j}^{m,n} = \left\{ d_{i,i}^{m,m-1}, d_{i,i}^{m,m+1},
        d_{j,j}^{n,n-1}, d_{j,j}^{n,n+1},
    2 {\tau_{\text{max}}} \right\}, 
\end{equation}

\noindent the dynamical delay, $\tau_{i,j}^{m,n}$, is defined as half of the minimal waiting time of subsequent events in both time series around event $e_i^m$  and $e_j^n$  and not larger than a predefined maximal value $\tau_{\text{max}}$ (Fig.~\ref{fig:network_construction}\,b), 

\begin{equation}
    \tau_{ij}^{m,n} = \frac{1}{2} \min_{\forall d \in D_{ij}^{m,n}} d \; .
    \label{eq:tau_mn}
\end{equation}

It encodes a small deviation between the occurrences, allowing for a time delay between two events. 
The parameter $\tau_{\text{max}}$ separates timescales of ERE synchronization and is set to a maximum delay of $\tau_{\text{max}} =10$ days (note that the results also remained stable for smaller and larger choices of $\tau_{\text{max}}$). 
The event synchronization strength $Q_{i,j}$ between locations $i$ and $j$ is the sum of all synchronous time points between all pairs of event sequences $\{t^m_i\}_{m=1}^{s_i}$ and $\{t^n_j\}_{n=1}^{s_j}$, 

\begin{align}
    Q_{i,j} &= \sum_{m=1}^{s_i} \sum_{n=1}^{s_j} S_{i,j}^{m,n} \hspace{1cm} \text{where} \hspace{0.5cm}
    S_{i,j}^{m,n} = \begin{cases}
    1&  \hspace{.5cm} 0< d_{ij}^{m,n} < \tau_{i,j}^{m,n} \;, \\
    0&  \hspace{.5cm}  \text{otherwise}\;.
\end{cases}
\end{align}

Blocks of consecutive events are counted as one event, placed on the point in time of the first event to avoid the dynamical delay $\tau_{i,j}^{m,n}$ resulting in a value of $1/2$, leading to a case where two sequentially occurring events would not be recognized as synchronous.

\paragraph{Statistical significance test.} We estimate the adjacency matrix $\mathbf{A}$ (Fig.~\ref{fig:network_construction}\,c) using a null-model test. 
Our null hypothesis is that an observed $Q_{i,j}$ value occurs from a pair of purely random event sequences with the same number of events $s_i, s_j$ as in the observed sequences. To encode the null hypothesis, we construct surrogate event sequences $e_i', e_j'$ with $s_i$, $s_j$ randomly uniformly distributed events.  
Event series $e_i$ is considered to be significantly synchronous to $e_j$ if their corresponding $Q_{i,j}$ value is higher than the $95$ percentile of $Q_{i', j'}$ values obtained using  $2000$ pairs of surrogate event sequences $e_i', e_j'$. 
Significant $Q_{i,j}$ values imply that we place an edge from node $n_i$ to $n_j$.

\paragraph{Link bundle correction.} As the number of comparisons becomes very high even for moderately large datasets (in our case $10^8$ comparisons), there is a non-negligible chance to consider singular pairs of time series as statistically significant, even though their significance is just by coincidence \citep{Haas2022}. 
To avoid such spurious links, we assume that synchronous time series are supposed to be caused by physical mechanisms and thus show spatially coherent patterns \citep{Boers2019}.  
For each spatial location, we rewire $2000$ times its network links randomly. 
We use a Gaussian kernel density estimator (KDE) with the bandwidth selected according to Scott's Rule of Thumb to compute the spatial link distribution of every random sample. A link is only found significant if its regional link distribution (also obtained by a Gaussian KDE) is above the $99.9$ percentile.

\subsection{Estimating communities within climate networks}  \label{sec:community_detection}
Determining macroscale regions of synchronously occurring EREs translates to identifying communities within the network.  
We contend that the Stochastic Block Model (SBM) is a suitable method for this purpose because: (i) it is not limited to assortative structures with more links inside the groups than between groups, and (ii) it allows us to describe both the relationship within and between the inferred groups.   
We use the Bayesian SBM algorithm  developed by  \cite{Peixoto2014} which offers several advantages: (i) the depth of the hierarchy as well as the number of groups in a layer can be inferred automatically or set as predefined, (ii) it avoids over- and underfitting, (iii) its implementation, a C++ based python module called \texttt{graph\_tool}, is computationally very fast, making it feasible to analyze large climate networks \cite{Peixoto2014}.

The algorithm uses an agglomerative multilevel Markov chain Monte Carlo (MCMC), thus, the inference algorithm is stochastic and returns a different number of groups per level in each run. 
Therefore, it is not guaranteed that a single run of the inference algorithm will converge to the best solution \cite{Peixoto2014}. 
However, as our network contains an inductive bias by its construction from stochastic EREs, multiple runs serve as an indicator of structural uncertainties in assigning a node to a community of nodes.   
We use a simple heuristic to estimate the posterior likelihood that a geographical location belongs to a particular climate network community.   
The number of communities that describes best the network structure is an outcome of the algorithm.
As we are interested in macroscale structures, we restrict the algorithm to contain $10$ communities at most.   
We run the SBM algorithm $100$ times. Most of the runs identify $ 6$ communities (with very few exceptions of $5$ and $7$ communities) with similar spatial shapes  (Fig.~\ref{fig:network_construction}\,d).   
Next, the overlap of all $100$ SBM runs allows us to estimate the likelihood of each spatial location belonging to each community. 
We call this the ``membership likelihood'' and define that locations within a community are those that have a membership likelihood larger than $90\,\%$. The membership likelihood also provides an estimate of the structural uncertainties involved (Fig.~\ref{fig:msl_all}).

\subsection{Estimating regions of synchronous EREs} \label{sec:sync_index}
We identify specific points in time of high synchronization between the sets of location $A$ and $B$ by a time series $t_{A\rightarrow B}^m$. 
For each time step $m$ the time series denotes the number of events in the region $A$ with a subsequently associated event to any other time series within $B$, expressed as
\begin{align}
    t_{A\rightarrow B}^m := \vert\vert   \left\{(i,j) \in A\times B :  -\tau_{ij}^{m,n}<d_{ij}^{m,n}<0 \wedge \vert d_{ij}^{m,n} \vert < \tau_{\text{max}} \right\} \vert\vert  \; ,\label{eq:points_high_sync}
\end{align}
where $\vert \cdot \vert$ denotes the absolute value, and $\vert \vert \cdot \vert \vert$ set cardinality.   
For the special case, $A=B$ we identify points in time of high synchronization \textit{within} a community. We label this time series as ``synchronous ERE index'' for community $A$. 
We define the points of exceptionally strong synchronization as these local maxima of the time series $t_{A\rightarrow A}$ that are also above the 90th percentile and call these ``days of maximum synchronization''.

\subsection{Estimation of conditional probabilities} \label{sec:cond_indpendence_test}
The probability for the occurrence of synchronous rainfall days within a cluster (denoted as $s=1$) under a condition $a$ is calculated as follows: 
Assume the set of days with exceptionally synchronous events being $S$ and the set of days that fulfill the condition $a$ being $A$. 
Then $ P(s=1|a) = \frac{P(s=1, a)}{P(a)} = \frac{||S \cap A||}{||A||}$ describes the conditional probability for synchronous events under condition $a$ . 
Here, $||\cdot||$ denotes set cardinality and $S\cap A$ the intersection of $S$ and $A$. 
Accordingly, the conditional probability for a second condition $b$ with a set of days $B$ is computed as:
\begin{equation}
    P(s=1|a, b) = \frac{P(s=1, a)}{P(a, b)} = \frac{||S \cap A \cap B||}{||A\cap B||} \, .  \label{eq:cond_probs}
\end{equation}
A corresponding null model is estimated by counting the days of maximum synchronization $||S||$ divided by the total number of days (i.e. in our case $\leq 0.1$). 
Hence, the upper limit for the null model is $P_{\mathrm{null model}}(s=1) = 0.1$. 

\subsection{Clustering of propagation times} \label{sec:clustering_prop_times}
The BSISO propagation events we define as a day of maximum synchronization within the region EIO (Fig.~\ref{fig:community_detection}\,a and Sec.\,\ref{sec:sync_index}) and are denoted as day $0$. 
Consecutive dates by less than $20$ days are removed. 
In total $110$ events are considered. 
We choose the propagation time range to be $5$ days before and $30$ days after the initiation day $0$. 
The propagation of different synchronous extreme rainfall events is investigated by a $K$-means cluster analysis \citep{Lloyd1982}. 
To account for the eastward as well as the northward propagation, propagation patterns are analyzed by Hovmöller diagrams of the OLR anomalies along a zonal band averaged between $0\degree{S}$ and $10\degree{N}$ and a meridional band averaged between $70\degree{E}$ and $80\degree{E}$.
OLR is used instead of precipitation because it directly indicates deep convection in the tropics \citep{Waliser1993}.  
To ignore daily variations and to make the macroscale propagation patterns better distinguishable, we apply a 2D-smoothing Gaussian filter on the Hovmöller diagrams with 5 Pixels as the width of the filter. 
We use the silhouette coefficient method to determine the optimal number of groups and find that the samples can be best fitted into $3$ distinct clusters.  
The silhouette coefficient indicates how similar a member is to its own cluster. 
We use it properly to remove outliers that have a silhouette coefficient lower than $0.05$ from the cluster analysis. 
This further reduces the number of input samples by $13$ events to $97$ events in total. 

\subsection{Local overturning circulation analysis} \label{sec:overturning_circulation}
In order to assess the relative contributions of the mass fluxes in the troposphere to the pair of meridional and zonal overturning circulations we use the method by \cite{Schwendike2014, Hu2017, Raiter2020, Galanti2022}. 
A Helmholtz decomposition is applied to the wind field $\textbf{V}=(u,v)$, to separate the divergent component from the rotational component as $\textbf{V} = \textbf{V}_{\mathrm{div}} + \textbf{V}_{\mathrm{rot}}$. 
The meridional (zonal) component of the divergent wind $\textbf{V}_{\mathrm{div}} = (u_{\mathrm{div}}, v_{\mathrm{div}})$ is associated with the north-south (east-west) oriented circulations, commonly known as the Hadley (Walker) cell.
The longitudinally dependent meridional circulation is calculated as the mass streamfunction $\Psi_v$ as a vertical integration over the pressure levels:
\begin{equation}
    \Psi_v(\lambda, \phi, p, t) =  \frac{2\pi R }{g} cos(\phi) \int_0^p \mathrm{d}p' v_{\mathrm{div}}(\lambda, \phi, p', t) \; ,
\end{equation}
where $R$ denotes Earth's radius, $g$ the gravitational constant, $\phi$latitude, $\lambda$longitude, $p$pressure level and $t$ time. 
The zonal mass streamfunction $\Psi_u$ is analogously computed as:
\begin{equation}
    \Psi_u(\lambda, \phi, p, t) =  \frac{2\pi R }{g} \int_0^p \mathrm{d}p' u_{\mathrm{div}}(\lambda, \phi, p', t) \; ,
\end{equation}
In our analysis, we use a simplified representation of $\Psi_u$ and $\Psi_v$ by averaging between $400$--$600$\,hPa.

\subsection{Vertical shear mechanism} \label{sec:vertical_shear}
One proposed mechanism for the northward propagation of the BSISO is the vertical shear mechanism. It suggests that the easterly negative vertical shear of the zonal mean flow (i.e. low-level westerlies and upper-level easterlies) generates a northward component to the moist Rossby waves \citep{Wang1997, Wang2005, Hoskins2006}. 
As the upward motion $w$ induced by the convective activity in the Indian Ocean decreases northward, the interaction of the BSISO with the easterly vertical shear of the mean flow $\Bar{U}_z$ generates cyclonic vorticity $\zeta_B$ and boundary layer convergence to the north of the anomalous convection band that promotes the northward movement. 
This process is described by a two-dimensional barotropic vorticity equation \citep{Wang1997}:
\begin{equation}
    \frac{\partial\zeta_B}{\partial t} \propto \Bar{U}_z  \frac{\partial w}{\partial y} \; . \label{eq:northward_prop_vorticity}
\end{equation}


\section*{Open Research}
All data needed to evaluate the conclusions in the paper are present in the paper or the Supplementary Materials.
Datasets for this research are available from Copernicus Climate Change Service and the MSWEP dataset \citep{Beck2019}. 
The data for the composite analysis from 1979 till date was taken from \cite{ERA5}. 
The code for generating and analyzing the networks is made publicly available under \cite{CodeClimnet}. 
The code for reproducing the analysis of the network communities and the spatial clustering described in this paper is publicly available under \cite{CodeCommunities}.
\section*{Author's contribution}
F.S. and B.G. conceived and designed the study. F.S. conducted the analysis. F.S. and B.G. prepared the manuscript. All authors discussed the results and edited the manuscript. 
\section*{Acknowledgements}
The authors declare that they have no competing interests.
F.S., J.S., and B.G. acknowledge funding by the Deutsche Forschungsgemeinschaft (DFG, German Research Foundation) under Germany’s Excellence Strategy – EXC number 2064/1 – Project number 390727645. F.S. and J.S. thank the International Max Planck Research School for Intelligent Systems (IMPRS-IS) for supporting their PhD program. N.B. acknowledges funding by the Volkswagen foundation, the European Union's Horizon 2020 research and innovation program under grant agreement No 820970 and under the Marie Sklodowska-Curie grant agreement No. 956170, as well as from the Federal Ministry of Education and Research under grant No. 01LS2001A.

\bibliography{./library.bib}

\newpage


\appendix
\renewcommand{\thefigure}{S\arabic{figure}}
\setcounter{figure}{0}
\setcounter{section}{0}

\section*{Supplementary Information}
\section*{Contents of this file}
\begin{figure}[!tb]
    \centering
    \includegraphics[width=.7\linewidth]{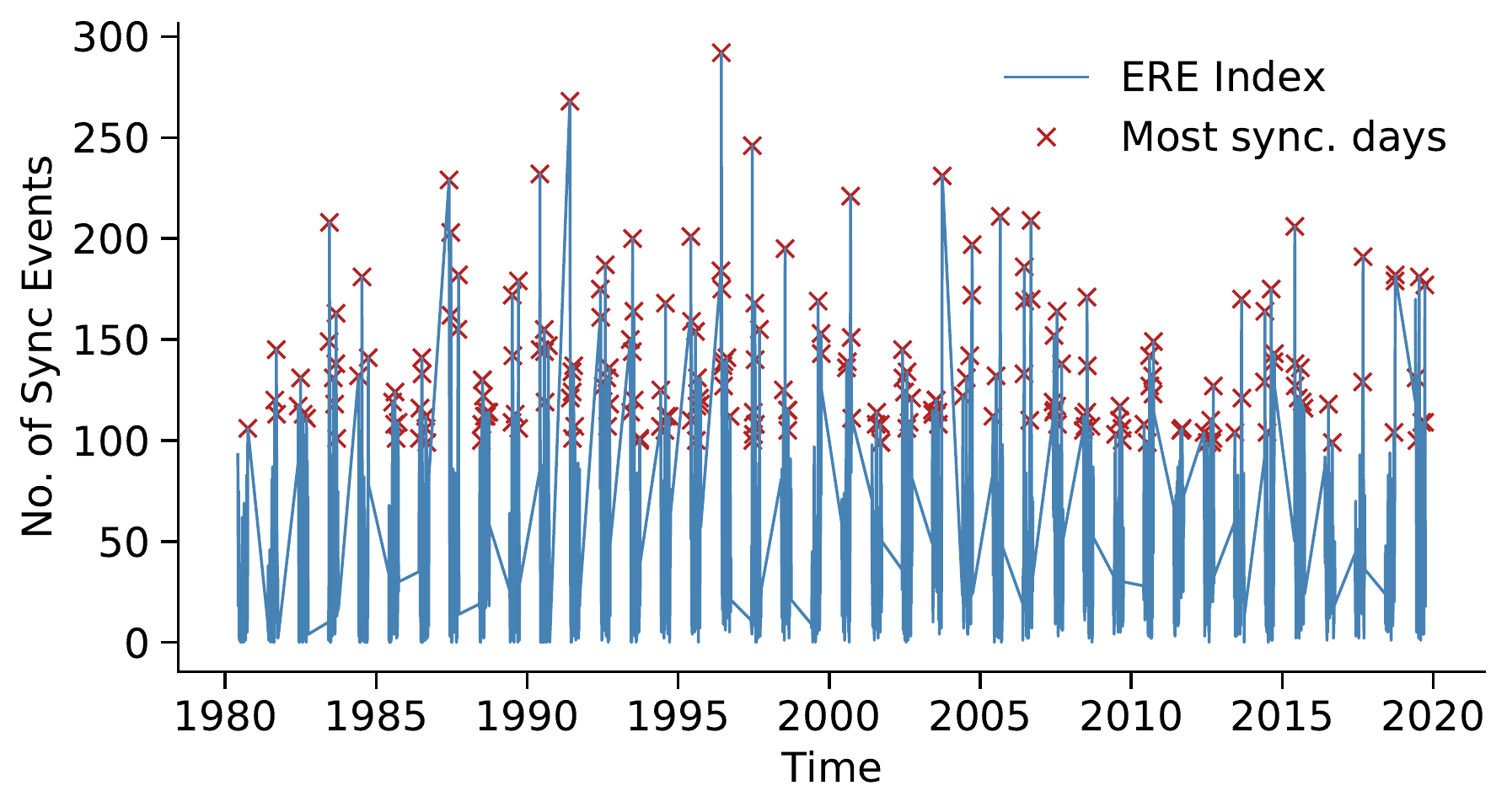}
    \caption{\textbf{Definition of synchronous extreme rainfall event index.} The synchronous extreme rainfall event index (ERE index) (eq.\,\ref{eq:points_high_sync}) is shown here as exemplary for the community in the equatorial Indian Ocean (Fig.~\ref{fig:community_detection}). We derive the most synchronous days by taking the peaks of all values above the $90$th percentile denoted by red crosses.  
	}
	\label{fig:ll_all_to_all}
\end{figure}

\begin{enumerate}
\item Text \ref{si:ere_def} to \ref{si:china_india_connection}
\item Figures \ref{fig:ee_q_map} to \ref{fig:dayprogression_olr_ch}
\end{enumerate}


\section*{Introduction}
\begin{figure}[!tb]
    \centering
    \includegraphics[width=1.\linewidth]{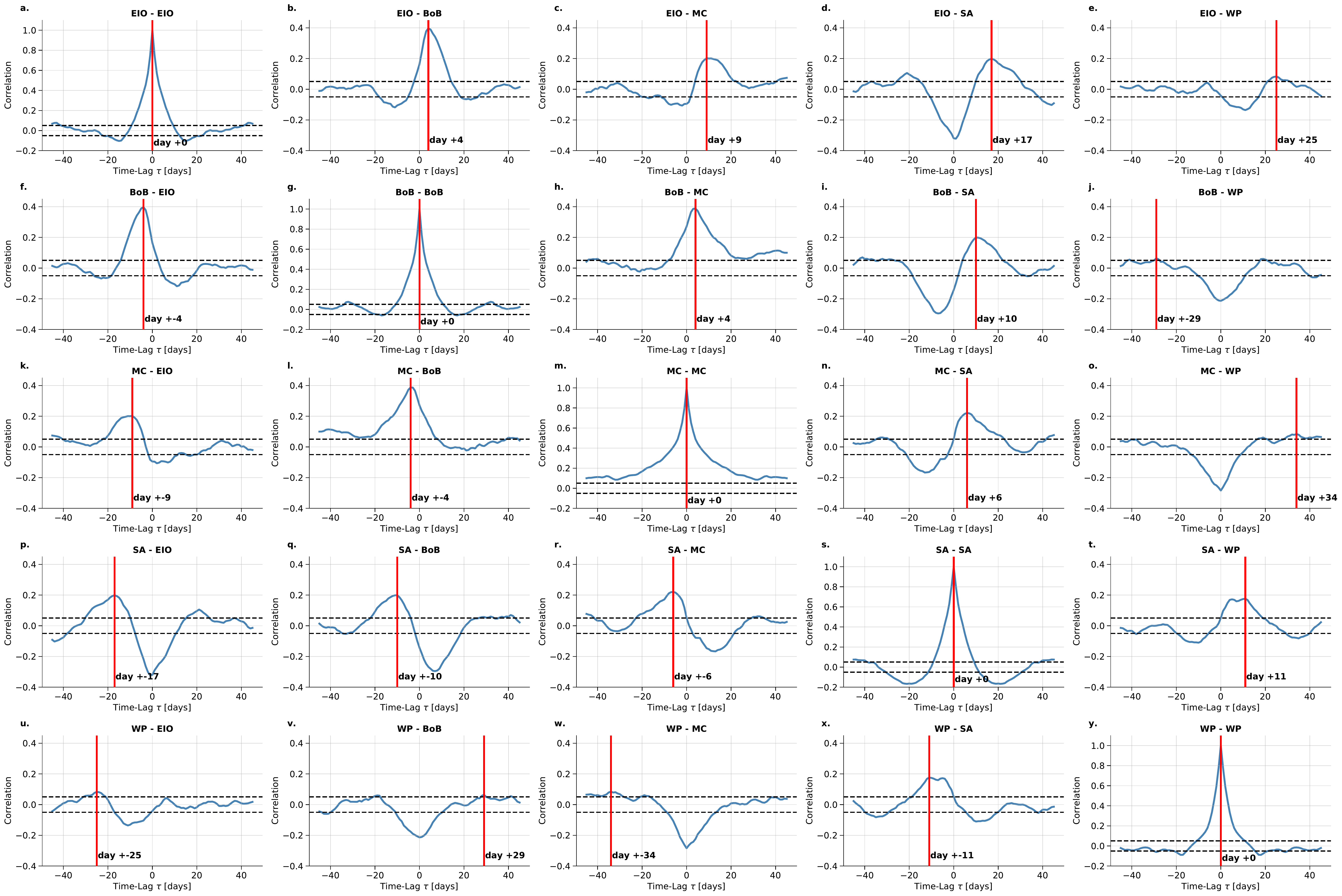}
    \caption{\textbf{Lead-Lag analysis for different communities.} Using the synchronous rainfall index (eq.\,\ref{eq:points_high_sync}) we derive lead-lag correlation analysis based on Spearman's rank order correlation. The dashed line denotes the 0.95 confidence interval. Vertical solid red lines denoting the day of maximum correlation.  
	}
	\label{fig:sync_ere_index}
\end{figure}

In this Supplementary Material to our article, we describe in detail the definition of Extreme Rainfall Events (EREs) during South Asian Summer Monsoon (SASM) (Text \ref{si:ere_def}, Fig.~\ref{fig:ee_q_map}). 
The schematic of the Event Synchronization approach is discussed in \ref{si:event_sync} and Fig.~\ref{fig:event_sync} 
A detailed analysis of the regions of synchronous EREs, detected by a community detection approach is given in \ref{si:msl_communities}. 
Here, we discuss as well the synchronous ERE index estimation per community (Fig.~\ref{fig:msl_all},\ref{fig:sync_ere_index}). 
In section \ref{si:propagation_BSISO} we show the spatially resolved propagation of the BSISO for the three discovered BSISO propagation modes (Fig.~\ref{fig:dayprogression_pr_canonical}, Fig.~\ref{fig:dayprogression_pr_eastward_blocked}, Fig.~\ref{fig:dayprogression_pr_stationary}). 
We demonstrate that the organization of EREs through the BSISO can also be shown by using a simple Linear Regression Model (Sec.~\ref{si:linear_model})
We provide a further discussion on the MJO relation in Sec.~\ref{si:mjo_connection} (Fig.~\ref{fig:mjo_phases}). 
More detailed information to the connection the El Ni\~no Southern Oscillation (ENSO) is given in Sec.~\ref{si:enso_connection} (Fig.~\ref{fig:nino34_propagation_modes}, Fig.~\ref{fig:vertical_cuts_enso}). 
In the main text we focused on the BSISO-dominated communities.
A detailed discussion on the North India China region and its connection to current literature is provided in Sec.~\ref{si:china_india_connection} (Fig.~\ref{fig:dayprogression_v_ch}, Fig.~\ref{fig:dayprogression_vimd_ch}, Fig.~\ref{fig:dayprogression_olr_ch}). 

\section{Extreme Rainfalls during the South Asian Monsoon} \label{si:ere_def}
\begin{figure}[!htbp]
	\centering
	\includegraphics[width=1.\textwidth]{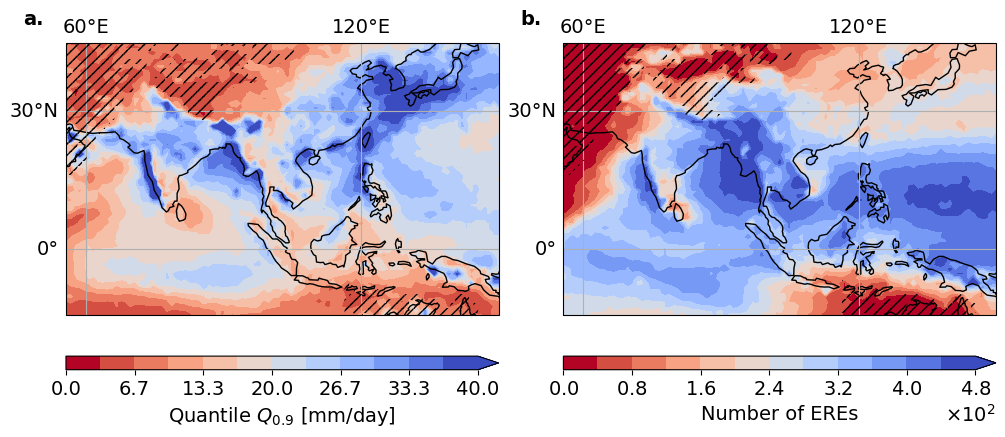}
	\caption{\textbf{Quantiles and total numbers of extreme-rainfall events.} \textbf{a} yields the number of counted EREs per spatial location. \textbf{b} shows the respective values of the quantile local function $Q_{0.9}$ for each location.
    Data is used from the MSWEP dataset \citep{Beck2019} over the period from 1979–2021.
    An Extreme Rainfall Event (ERE) is defined as a day with more than 1 mm/day of precipitation. Locations with less than $10$ events in total are excluded from the analysis and marked as hatched areas.}
	\label{fig:ee_q_map}
\end{figure}

To construct the climate network of EREs in the South Asian Monsoon region, we only take into account those grid locations where the 95th percentile value of rainfall is greater than 10 mm/day or if we find more than $ 10 $ such `event days' over the whole time period. 
Regions around the Arabian Peninsula and the Gobi and Taklamakan deserts were thus excluded from the analysis (compare Figure~\ref{fig:ee_q_map}). 
We find that number of extreme events varies significantly between the tropics and the subtropics, with the tropics experiencing a far higher number of events (Figure~\ref{fig:ee_q_map}).  
Our null model (sec.\,\ref{sec:es_nets}) which depends on the number of events in each time series, helps to reduce the bias introduced in the event synchronization measure due to different event rates \citep{Rheinwalt2016}. 
The upper bound on the dynamical delay is set to $\tau_{\text{max}}=10$ days for all grid location pairs. 

\section{Event Synchronization} \label{si:event_sync}
\begin{figure}[!h]
	\centering
	\includegraphics[width=.7\textwidth]{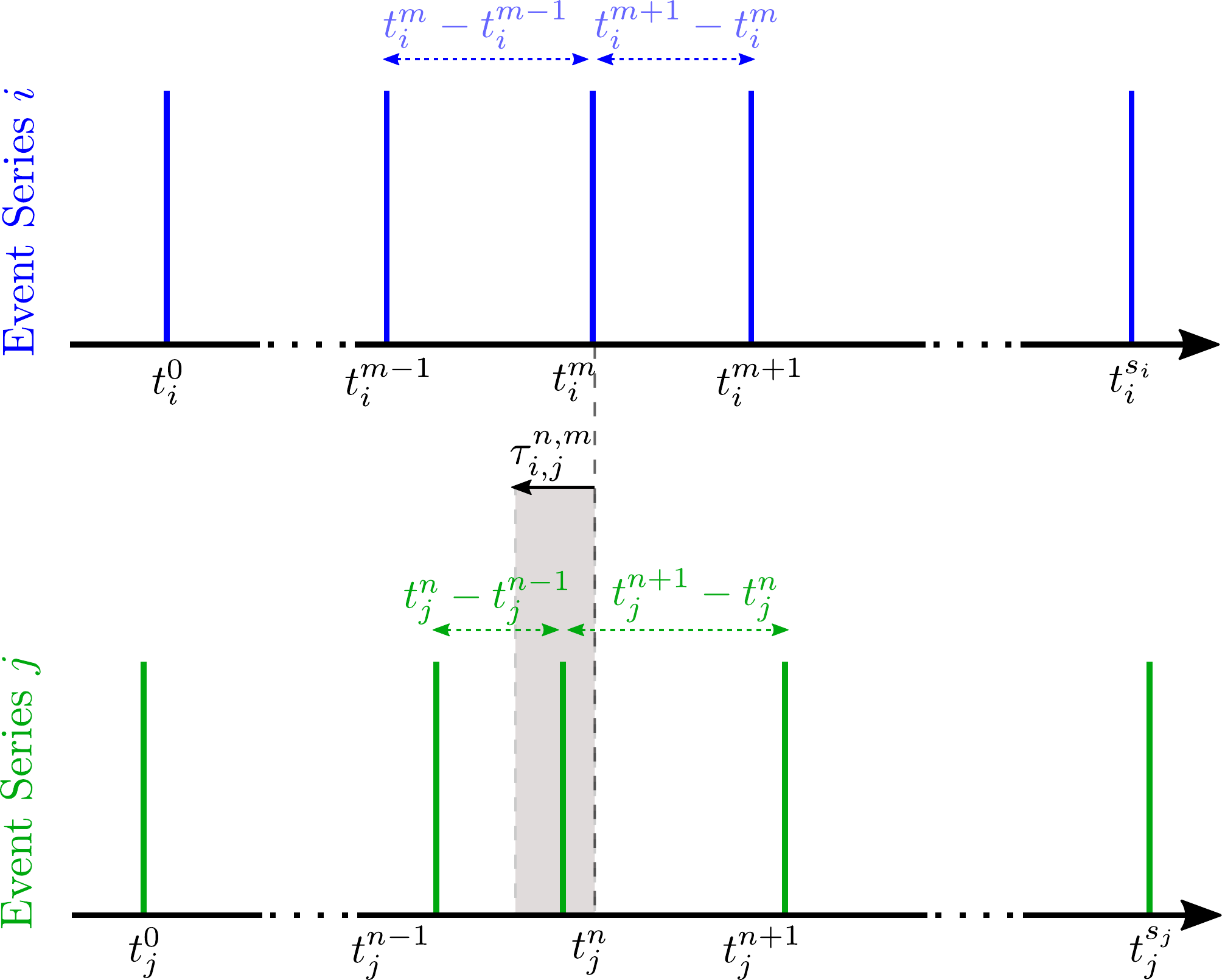}
	\caption{\textbf{Scheme of time-delayed Event Synchronization.} The point in time  $ t_i^m $ is identified as synchronous to $ t_j^n $ since the time difference between $ t_i^m $ and $ t_j^n $ is within the allowed the range given by $\tau_{ij}^{m,n}$ according to the definition expressed in equation (\ref{eq:tau_mn}). Not shown here is the maximum delay $\tau_{\text{max}} $. }
	\label{fig:event_sync}
\end{figure}

Fig.~\ref{fig:event_sync} describes the ES scheme. Note that if two ore more events in one series occur at subsequent time steps the dynamical delay $\tau_{i,j}^{m,n}$ will result in a value of 1/2, leading to a case where it is likely that two sequentially occurring events are not recognized as synchronous. Therefore, in our analysis, blocks of consecutive events are counted as one event, placed on the point in time of the first event.

\section{BSISO drives the organization of synchronous EREs}  \label{si:msl_communities}
\begin{figure}[!tb]
    \centering
    \includegraphics[width=1.\linewidth]{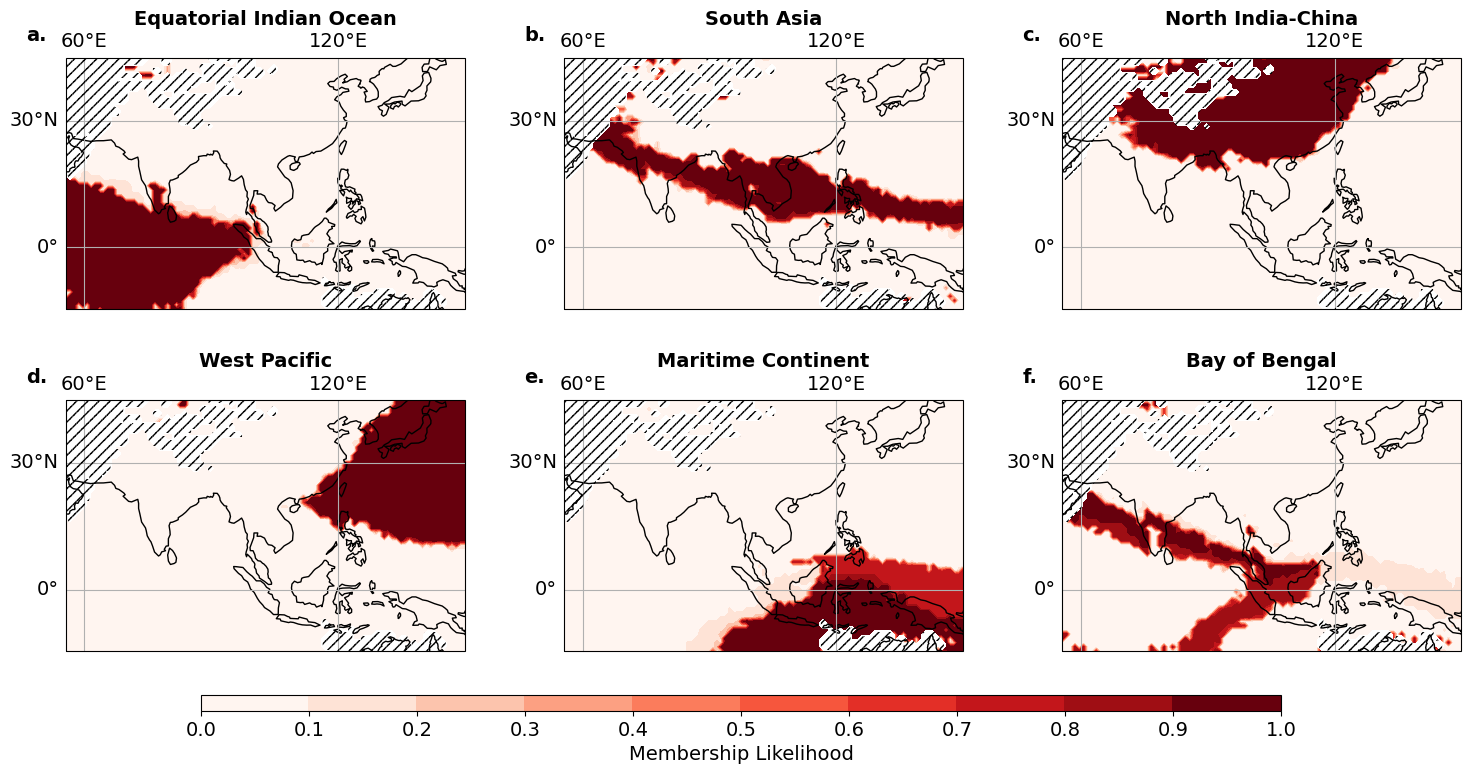}
    \caption{\textbf{Membership Likelihoods for different communities.} Using the heuristic outlined in sec. \ref{sec:community_detection} we find 6 stable communities. The colorbar shows the membership likelihood of a respective community. $100$ independent runs of the community detection algorithm have been used for this analysis. 
	}
	\label{fig:msl_all}
\end{figure}

We compute the membership likelihood for the 6 identified monsoon regions in Fig.~\ref{fig:community_detection}\,a. 
These are computed as it is described in Sec.\,\ref{sec:community_detection}. 
The membership likelihoods for the different climate network communities is visualized in Fig.~\ref{fig:msl_all}. 
These plots show low spatial variability of the communities and therefore underline the robustness of the community detection algorithm. 

The BSISO is directly reflected in the shape of the communities. The BSISO is known to emerge around the center of the equatorial Indian Ocean characterized by intense rainfalls \citep{Lau2011, Kim2018} consistent with our observation of an increased likelihood of synchronous EREs in the EIO community for BSISO phases 1 and 2 (Fig.~\ref{fig:bsiso_phases}\,a) \citep{Kikuchi2021}.   
The BSISO-driven rainfalls are known to intensify significantly over the Bay of Bengal \citep{Mishra2017} consistent with our observations (Fig.~\ref{fig:bsiso_phases}\, c). 
The increased probability of experiencing extreme precipitation during active BSISO by a factor of two to three in the MC community region is consistent with observations in other studies \citep{DaSilva2021}.  
The strongest enhancement in SA occurs during BSISO phases 4 to 6 (Fig.~\ref{fig:bsiso_phases}\,d) establishing a monsoon trough developing here over the Bay of Bengal and leading to winds blowing southwesterly across the Indian Ocean carrying a lot of moisture. 
Many studies suggest that the establishment of a such convergence zone along the monsoon trough is enhanced during active BSISO expressed by strong low-pressure systems \citep{Mishra2017, Anandh2018, DiCapua2020a, Karmakar2021, Schreck2021, Hunt2022}.  
In WP, EREs occur during BSISO phases 6 and 7 (Fig.~\ref{fig:bsiso_phases}). 
The intraseasonal variation of the convection anomalies over the tropical eastern Indian Ocean until the western Pacific Oceans, so-called Pacific Japan (JP)-mode, has been attributed to the summer time MJO \citep{Li2019a}.


\section{Propagation pattern of EREs associated with BSISO} \label{si:propagation_BSISO}
\begin{figure}[!tb]
    \centering
    \includegraphics[width=1.\linewidth]{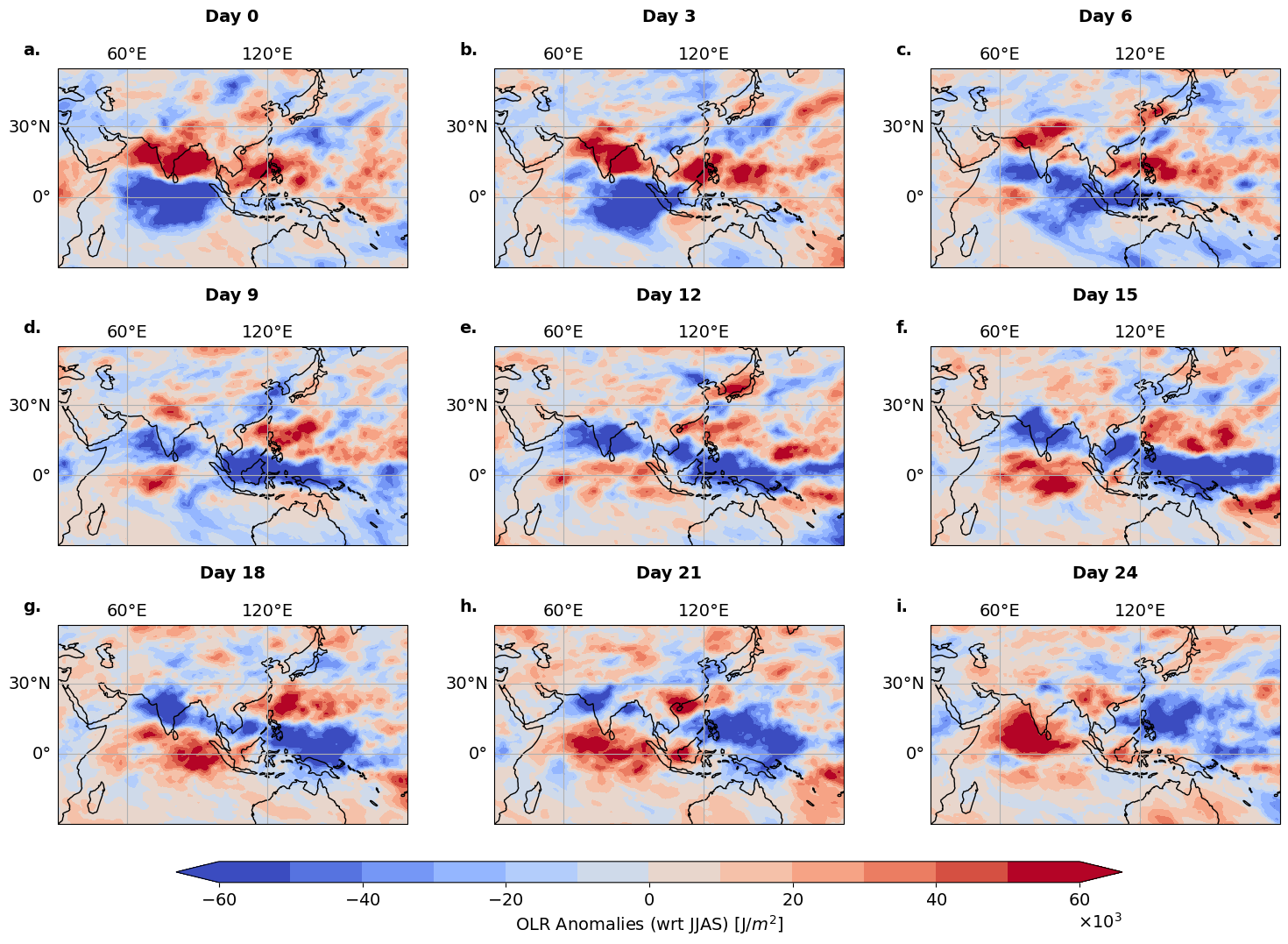}
    \caption{\textbf{Propagation of anomaly precipitation for Canonical BSISO propagation mode.} The rainfall anomalies are computed according to the JJAS climatology. Day $0$ denotes the days of maximum synchronization in the EIO community (Fig.~\ref{fig:community_detection}\,a) classified as ``Canonical'' BSISO propagation mode (Fig.~\ref{fig:propagation_times} 2nd row). The following plots \textbf{b}-\textbf{i} are the composites of the respective days after day $0$ in steps of $3$ days. 
	}
	\label{fig:dayprogression_pr_canonical}
\end{figure}

\begin{figure}[!tb]
    \centering
    \includegraphics[width=1.\linewidth]{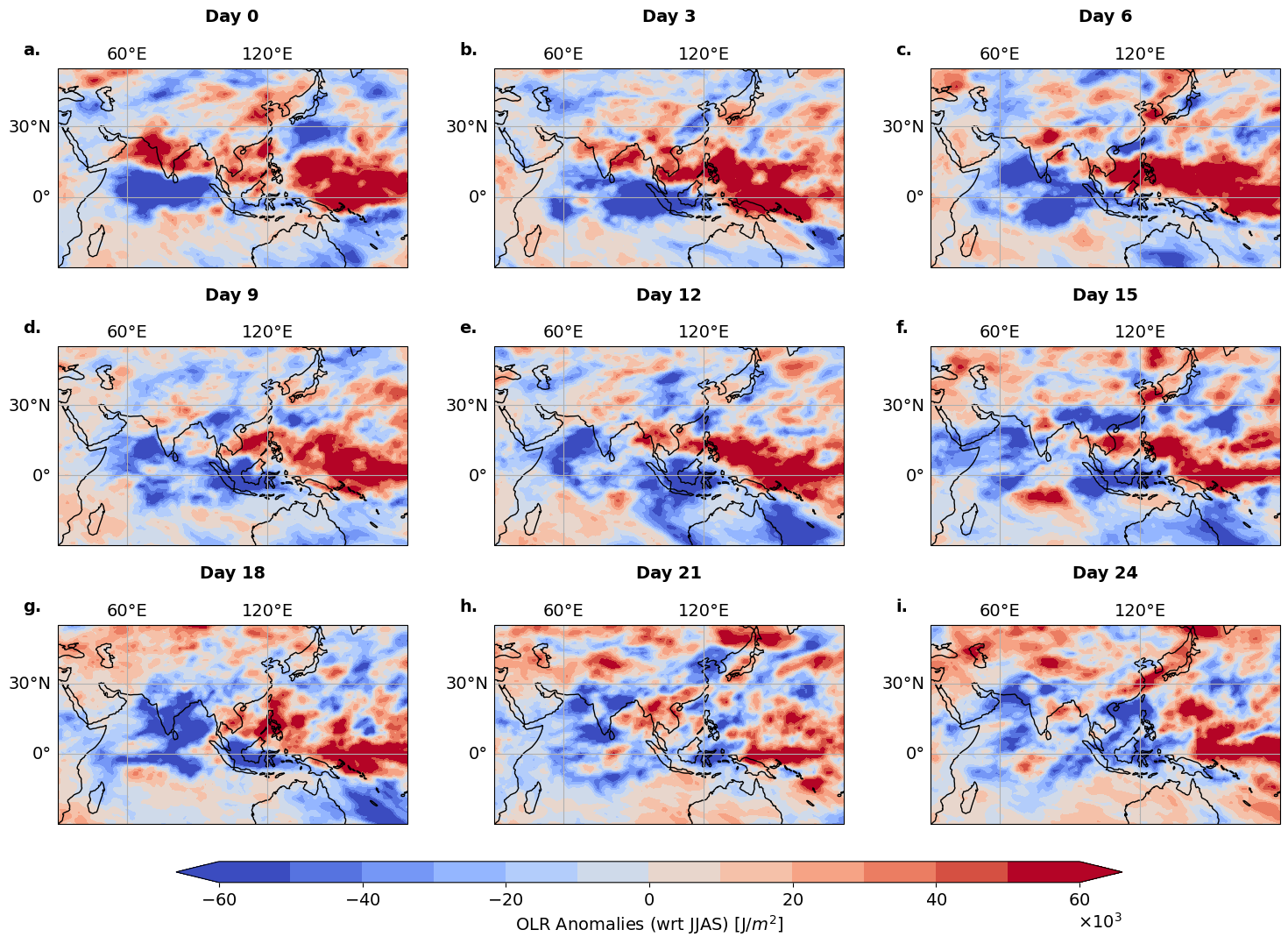}
    \caption{\textbf{Propagation of anomaly precipitation for Eastward Blocked BSISO propagation mode.} The rainfall anomalies are computed according to the JJAS climatology. Day $0$ denotes the days of maximum synchronization in the EIO community (Fig.~\ref{fig:community_detection}\,a) classified as ``slow'' BSISO propagation mode (Fig.~\ref{fig:propagation_times} 2nd row). The following plots \textbf{b}-\textbf{i} are the composites of the respective days after day $0$ in steps of $3$ days. 
	}
	\label{fig:dayprogression_pr_eastward_blocked}
\end{figure}

\begin{figure}[!tb]
    \centering
    \includegraphics[width=1.\linewidth]{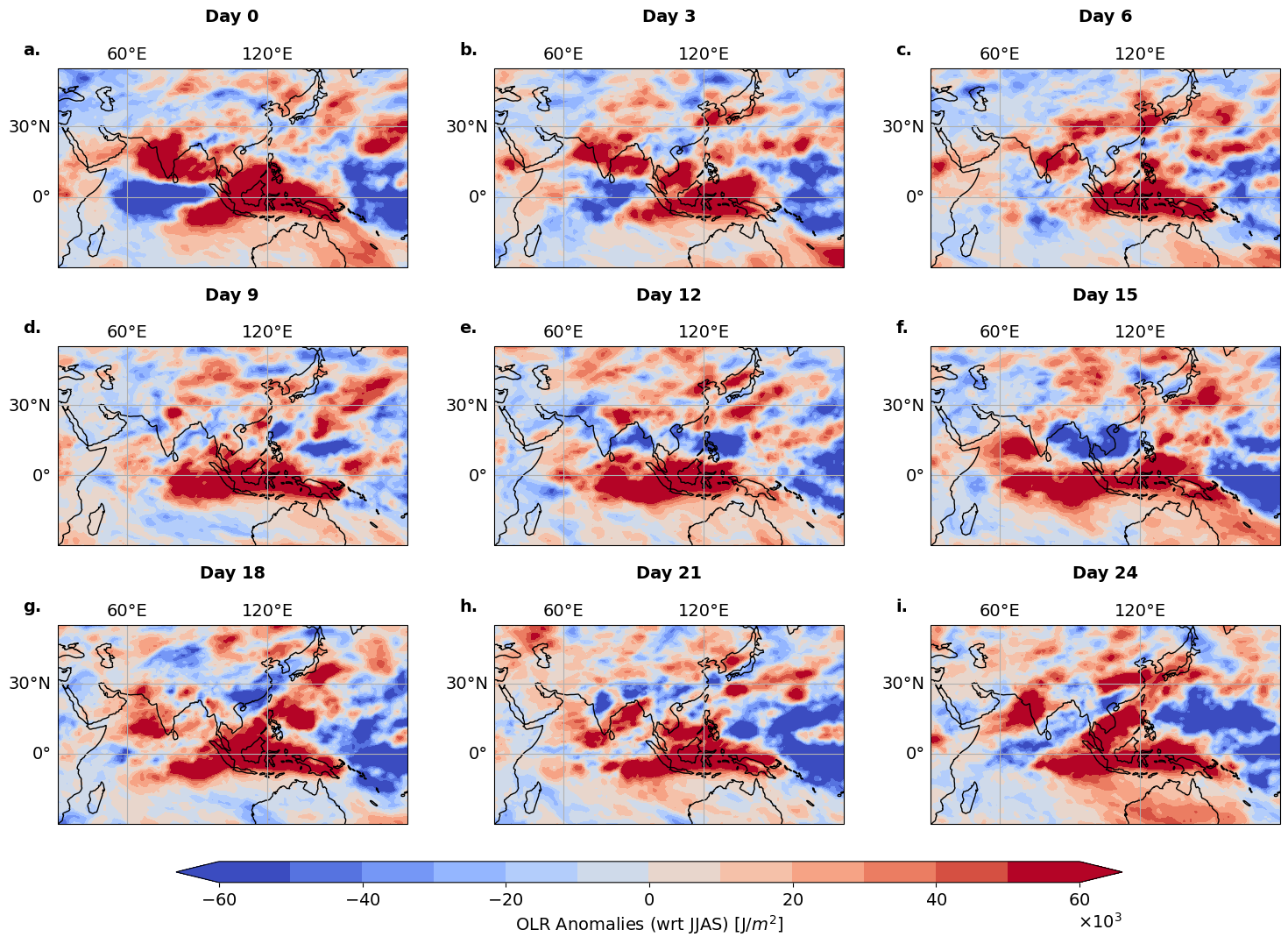}
    \caption{\textbf{Propagation of anomaly precipitation for Stationary BSISO propagation mode.} The rainfall anomalies are computed according to the JJAS climatology. Day $0$ denotes the days of maximum synchronization in the EIO community (Fig.~\ref{fig:community_detection}\,a) classified as ``suppressed'' BSISO propagation mode (Fig.~\ref{fig:propagation_times} 3rd row). Rainfall anomalies occur solely in the region of the EIO community. 
	}
	\label{fig:dayprogression_pr_stationary}
\end{figure}

The propagation of the BSISO is subdivided into 3 different modes. 
The propagation of the convective system is shown in a condensed way in Fig.~\ref{fig:dayprogression_all}. 
Here, we show spatial composites of all three propagation modes in their spatial progression until 24 days in advance. The Canonical propagation is shown in Fig.~\ref{fig:dayprogression_pr_canonical}, the Eastward Blocked mode in Fig.~\ref{fig:dayprogression_pr_eastward_blocked},  and the Stationary case in Fig.~\ref{fig:dayprogression_pr_stationary}. 

\section{Linear Model for BSISO index} \label{si:linear_model}
A target time series $Y(t)$ is modelled by $N$ input time series $X_i(t)$, where $i=1,\dots, N$. We use a simple multi-dimensional linear regression incorporating time lags $\tau_i$, expressed as:
\begin{equation}
    Y(t) = \sum_{i=0}^N a_i X_i(t-\tau_i) + b + \epsilon(t)\, ,  \label{eq:linear_model}
\end{equation}
where $a_0\dots a_N, b$ are the fitting parameters and $\epsilon(t)$ is the error term. 
The fitting parameters are estimated using the least square fit. The goodness of the fit (also denoted as ``explained variance'') is expressed by the square of the correlation between the observed $Y$ values and the predicted $\hat{Y}$ values as $r^2 = \sum ( \hat{Y}(t) - \overline{Y}(t) / Y(t)- \Bar{Y}(t) ) $, where $\overline{Y}$ denotes the time average. 

In order to quantify the amount of ERE variability that is explained by BSISO, we use the simple linear regression model between specific synchronous ERE indices (see Material\&Methods Sec.\,\ref{sec:sync_index}) and the BSISO1 and BSISO2 indices (see Material\&Methods Sec.\,\ref{sec:data}). 
We obtain the following explained variances: EIO $r=67\,\%$, BoB $r=77\,\%$, SA $r=78\,\%$, MC $r=54\,\%$, WP $r=70\,\%$ and NIC $r=8\,\%$. 
Conversely, BSISO variability is explainable from the community synchronization indices with explained variances of $r=83\,\%$ ($r=68\,\%$) for BSISO1 (BSISO2) indices.  
These results, obtained by using a simple linear regression (and applying a low-pass filter with 3 days cutoff on the time series to neglect small daily variations), affirm the close relationship of ERE synchronization to the BSISO.

\section{Connection to Madden Julian Oscillation} \label{si:mjo_connection}
\begin{figure}[!tb]
    \centering
    \includegraphics[width=1.\linewidth]{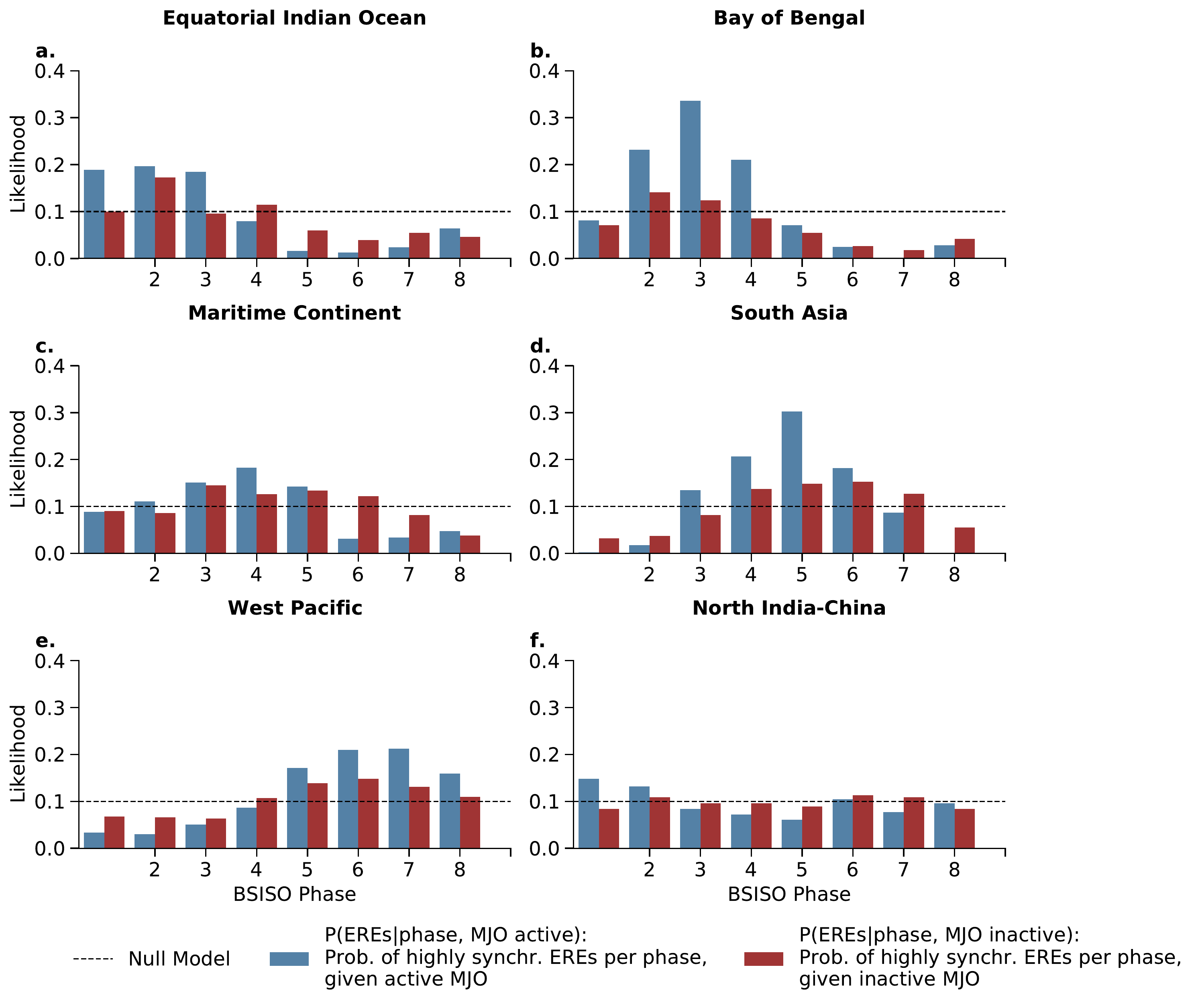}
    \caption{\textbf{Likelihood of synchronous events for active MJO phases}. The occurrence of synchronous events ($s=1$) is analyzed for active (break) MJO phases displayed as blue (red) bars.  The dashed line depicts the likelihood of synchronous events for a threshold of $10\,\%$ from a null model of a random distribution of synchronous events. \textbf{a-f} Likelihood of synchronous events $s=1$ for specific phases. The histograms are for the communities received from the analysis in Figure \ref{fig:community_detection}\, \textbf{d}, i.e.  
    \textbf{a} equatorial Indian Ocean, 
    \textbf{b} Bay of Bengal, 
    \textbf{c} Maritime continent, 
    \textbf{d} South Asia,
    \textbf{e} Western Pacific and
    \textbf{f} North India-China.
	}
	\label{fig:mjo_phases}
\end{figure}

The Madden Julian Oscillation (MJO) \citep{Madden1971} is a closely related phenomenon. 
Here, we show that the MJO alone is not sufficient to explain the organization of EREs during JJAS. 
To do so, we apply the same conditional independence test as for the BSISO using the RMM index as it was introduced by \cite{Wheeler2004}. 
The qualitative shape of the distribution remains, however, the lower likelihoods express that BSISO1 and BSISO2 indices are more suited for analysis of EREs during boreal summer (Fig.~\ref{fig:mjo_phases}).

\section{Comparison to classical ENSO conditions} \label{si:enso_connection}
\begin{figure}[!tb]
    \centering
    \includegraphics[width=1.\linewidth]{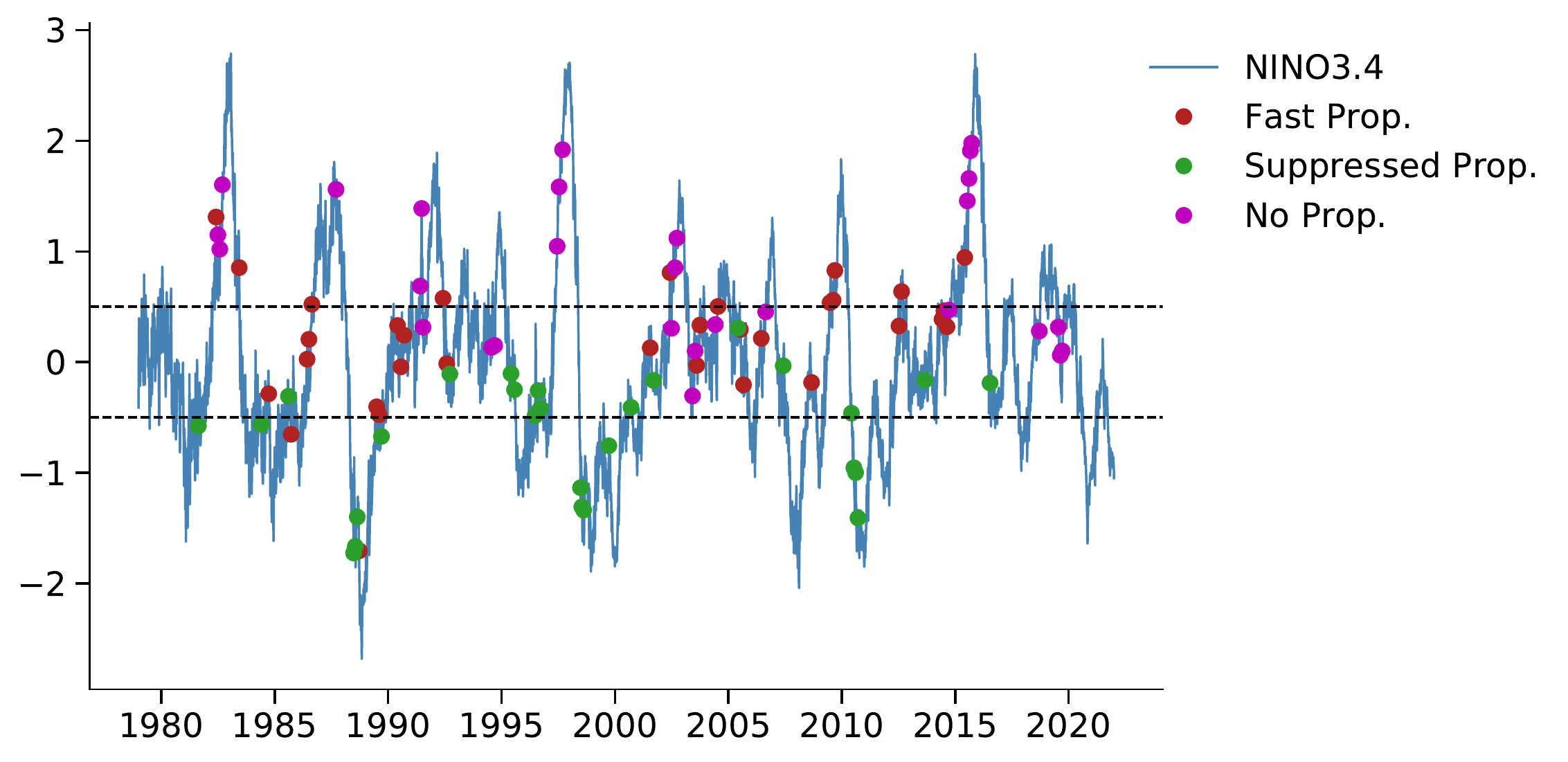}
    \caption{\textbf{propagation modes at NINO3.4 index}. The occurrence of synchronous events is displayed according to the identified propagation mode.  The solid line depicts the NINO3.4 index. The definition of an El Ni\~no (La Ni\~na) year was adapted from \citep{Capotondi2020}, represented by the dashed line of 0.5 as the respective threshold.
	}
	\label{fig:nino34_propagation_modes}
\end{figure}

We compare the single BSISO events to the respective ENSO state in the Pacific Ocean. Following \citep{Trenberth1997, Trenberth2001}, El Ni\~no and La Ni\~na are defined using the NINO3.4 index.  We use June to September daily SST anomalies and select El Ni\~no-like conditions (La Ni\~na-like conditions) based on the average JJAS SST anomalies of the NINO3.4 index region. The respective plot is shown in Fig.~\ref{fig:nino34_propagation_modes}.

\begin{figure}[!tb]
    \centering
    \includegraphics[width=.8\linewidth]{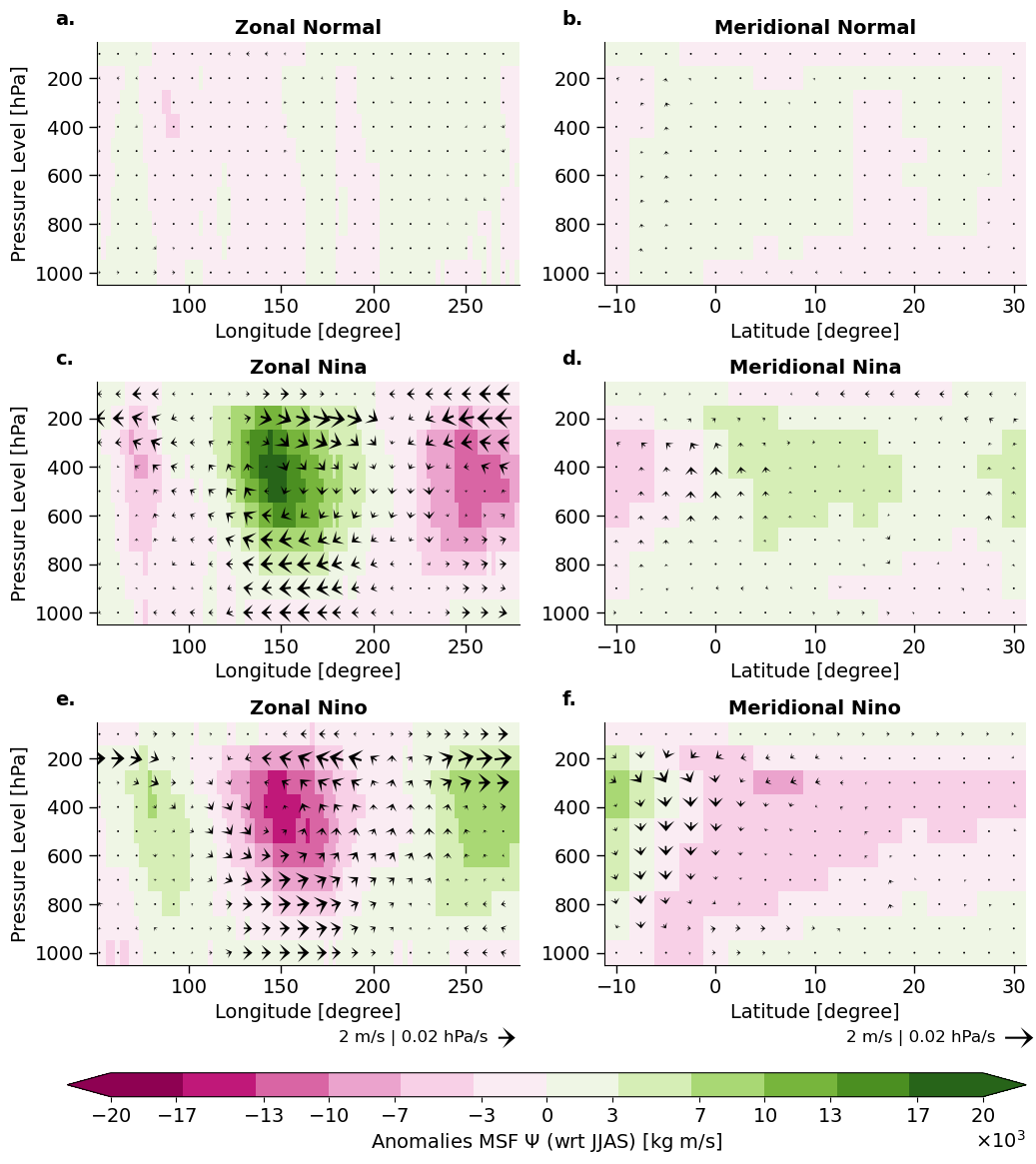}
    \caption{\textbf{Vertical cuts of overturning circulation for ENSO types.} Different ERE propagation modes are plotted for the background overturning circulation anomalies. 
    The first column (\textbf{a,c,e}) shows the composited vertical cuts in the zonal direction with the latitudes averaged between $10\degree{S}$-$10\degree{N}$, the second column (\textbf{b,d,f}) in the meridional direction in the Indian Ocean with the latitudes averaged between $60\degree{E}$-$120\degree{E}$. The color shading denotes the mass stream function in the zonal (meridional) direction.
    The wind fields are computed in the vertical direction from the u (v) components in m/s and in the horizontal direction in hPa/s.  
	}
	\label{fig:vertical_cuts_enso}
\end{figure}

We also analyze the classical circulation conditions for El Ni\~no-like conditions (La Ni\~na-like conditions) in the Pacific Ocean. The pressure level dependent plots are shown in Fig.~\ref{fig:vertical_cuts_enso}. 

\section{Connection of the rainfall extremes during Indian and East Asian Monsoon}  \label{si:china_india_connection}

One region that emerges from the community detection approach is the NIC community which is of all six regions the only one solely over land (Fig\,\ref{fig:community_detection}\,a). 
The community connects the northeastern part of India, the Tibetan Plateau, the Himalayan Mountains, and most parts of China of which the Himalayan foothills and the Ganges Delta in North East India are among those regions that experience the highest rainfall accumulation during SASM (Fig.~\ref{fig:ee_q_map}) during core monsoon season in July. 

\begin{figure}[!tb]
    \centering
    \includegraphics[width=.6\linewidth]{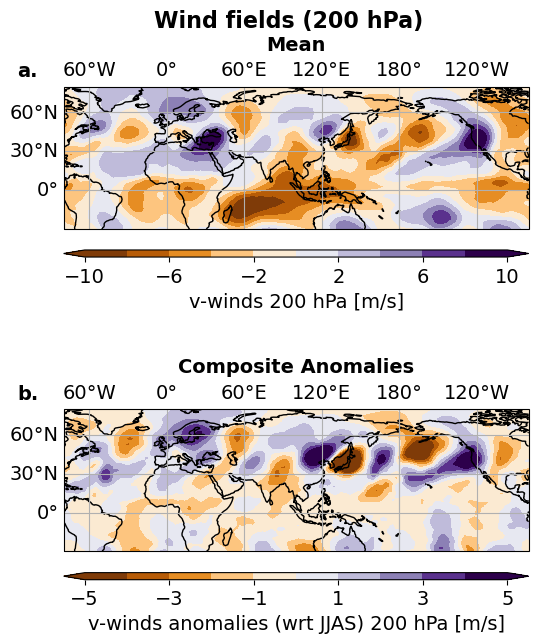}
    \caption{\textbf{Atmospheric conditions for the synchronous day pattern.} The v-winds components at $200$\,hPa is plotted for the mean (\textbf{a}) the composited anomalies \textbf{b}). The days were chosen as described in section \ref{sec:sync_index} using the days of maximum synchronization in the NIC region. The composite anomalies are computed with respect to the JJAS climatology. The Silk Road pattern connects clearly the North of India with the China area. 
	}
	\label{fig:dayprogression_v_ch}
\end{figure}

\begin{figure}[!tb]
    \centering
    \includegraphics[width=.6\linewidth]{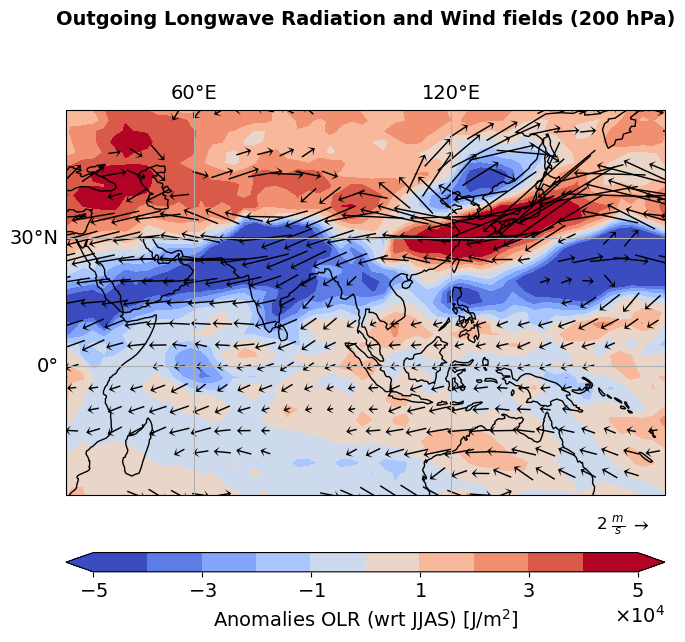}
    \caption{\textbf{Outgoing Longwave Radiation for days of maximum synchronization for NIC region} Same as Figure \ref{fig:dayprogression_v_ch} but for Outgoing Longwave Radiation (IVF) overlapped by wind fields at from anomalous upper-level wind fields at $200\,$hPa. Composites are computed for the days of maximum synchronization. The composite anomalies are computed with respect to the JJAS climatology. Only wind arrows that are significant at 95\,\% level following the Student's t-test are plotted. 
	}
	\label{fig:dayprogression_olr_ch}
\end{figure}

In early work, the enhanced upper-level atmospheric wave train, known as circumglobal teleconnection \citep{Ding2005} (or due to its near-equivalence over Eurasia \citep{Zhou2019} also known as the ``Silk Road pattern'' \citep{Enomoto2003}) was suspected to connect the northern Indian region with the China region \citep{Ding2005}, and \cite{Boers2019} suspect this mechanism to be responsible for the synchronization of EREs between northern India and northern China \citep{Boers2019}.  
Composite anomalies of the days of maximum synchronization within the NIC community reveal a large-scale wave train pattern originating from the mid-latitude Atlantic ocean.
It is enhanced across Eurasia and connects the northern India region with the Yellow River basin (Fig.~\ref{fig:dayprogression_v_ch}) corroborating results from \citep{Gupta2022}.  

\begin{figure}[!tb]
    \centering
    \includegraphics[width=.6\linewidth]{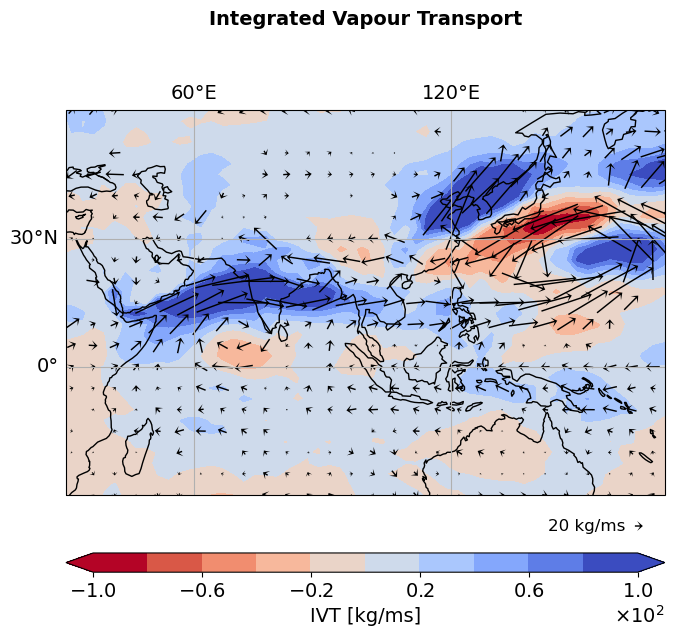}
    \caption{\textbf{Vertically integrated water vapor flux (IVT) for days of maximum synchronization for NIC region.} Same as Figure \ref{fig:dayprogression_v_ch} but for the vertically integrated water vapor flux (IVF). Day 0 denotes the days of maximum synchronization. The composite anomalies are computed with respect to the JJAS climatology. Only IVT arrows that are significant at 95\,\% level following the Student's t-test are plotted. 
	}
	\label{fig:dayprogression_vimd_ch}
\end{figure}

We also identify a further mechanism connecting parts of northern India with the Yellow River basin in northern China. Using composites of vertically integrated moisture vapor flux (IVF) uncovers a continuous path of anomalously high moisture transport established in a moisture corridor (Fig.~\ref{fig:dayprogression_vimd_ch}\,a-e), starting from the Ganges Delta and gated by the Tibetan Plateau towards the Yellow River Basin and northern China possibly driven by the Silk Road pattern. 
The connection manifests itself in the northward displacement of the western North Pacific subtropical high (WNPSH) during the Asian summer monsoon \citep{Pan2020} corroborating studies on the dominant route of stage 4 of the East Asian Atmospheric Rivers \citep{Pan2020}.

\clearpage
\newpage

\bibliography{./library.bib}


\end{document}